 \documentclass[11pt, reqno]{amsart}

\usepackage{amsaddr}

\usepackage[utf8x]{inputenc}
\usepackage{hyperref}

\usepackage{amsmath, amsthm, amsfonts, amssymb}
\usepackage{cite}
\usepackage{epsfig, enumerate}
\usepackage{color}
\usepackage{dsfont}

\usepackage{xfrac}

\usepackage{tikz}
\usetikzlibrary{patterns}

\usepackage{pdf14}

\usepackage[numbers,sort]{natbib}

\DeclareMathOperator{\rank}{{\rm rank}}

\theoremstyle{plain}

\newtheorem{theorem}{Theorem}

\newtheorem{proposition}{Proposition}
\newtheorem{lemma}[proposition]{Lemma}

\numberwithin{proposition}{section} 
\numberwithin{equation}{section} 

\theoremstyle{definition}

\newtheorem{definition}{Definition}

\newtheorem*{rem*}{Remark}

\newcommand{\lam}{E}

\newcommand{\norm}[1]{\left\| #1 \right\|}

\newcommand{\R}{\mathds{R}} 
\newcommand{\C}{\mathds{C}} 
\newcommand{\Z}{\mathds{Z}} 
\newcommand{\NN}{\mathds{N}} 

\newcommand{\M}{\mathds{M}}

\newcommand{\G}{\mathds{G}} 
\newcommand{\Hi}{\mathds{H}}  

\newcommand{\Sb}{{\mathbb S}}

\newcommand{\A}{\mathbb{A}}
\newcommand{\B}{\mathbb{B}}
\newcommand{\T}{\mathds T}

\newcommand{\D}{\mathds{D}}
\newcommand{\N}{{\mathbb N}}

\newcommand{\EE}{{\mathbb E}}
\newcommand{\VV}{{\mathds V}}

\newcommand{\Her}{\mathbf{Her}} 
\newcommand{\m}{\boldsymbol{\mu}} 
\newcommand{\n}{\boldsymbol{\nu}} 

\newcommand{\ot}{\leftarrow} 

\newcommand{\one}{{\bf I}}  
\newcommand{\nul}{{\bf 0}} 

\newcommand{\bff}{{\mathbf f}}
\newcommand{\bfh}{{\mathbf h}}
\newcommand{\bfu}{{\mathbf u}}
\newcommand{\bfv}{{\mathbf v}}
\newcommand{\bfw}{{\mathbf w}}

\newcommand{\Aa}{{\mathcal A}}
\newcommand{\Bb}{{\mathcal B}}

\newcommand{\Dd}{{\mathcal D}}
\newcommand{\Ll}{{\mathcal L}}
\newcommand{\Nn}{{\mathcal N}}
\newcommand{\Pp}{{\mathcal P}}

\newcommand{\bbb}{{\mathfrak b}} 

\newcommand{\qtx}[1]{\quad\text{#1}\quad}

\newcommand{\wh}{\widehat}

\newcommand{\pmat}[1]{\begin{pmatrix} #1  \end{pmatrix}}
\newcommand{\smat}[1]{\left( \begin{smallmatrix} #1  \end{smallmatrix} \right)}

\newcommand{\diag}{{\rm diag}}

\DeclareMathOperator{\Tr}{{\rm Tr}}

\DeclareMathOperator{\supp}{{\rm supp}}
\DeclareMathOperator{\spec}{{\rm spec}}

\newcommand{\ac}{{\rm ac}}

\newcommand{\dd}{{\rm d}}

\newcommand{\Ups}{\Upsilon}

\usepackage{hyperref}

\title[Multiplicity criteria for the ac spectrum]{Multiplicity Criteria for the absolutely continuous spectrum of finite hopping Hermitian Operators}

\author{Boris Berm\'udez-C\'ardenas$^1$, Christian Sadel$^{1,2}$, Amal Taarabt$^1$}
\address{$^1$Pontificia Universidad Cat\'olica de Chile, Facultad de Matem\'aticas,\\ Av. Vicuña Mackenna 4860, 7820436 Macul, Santiago, Chile} 
\address{$^2$Friedrich-Alexander-Universität Erlangen-Nürnberg, Department Mathematik, \\Cauerstr. 11, D-91058 Erlangen, Germany}
\email{bxbermudez@uc.cl, chsadel@uc.cl, amtaarabt@uc.cl}


\begin{document}

\begin{abstract}
We consider locally finite-hopping Hermitian operators that admit radial-type transfer matrix sets in the sense of  Sadel. We extend his spectral averaging formula from scalar spectral measures to matrix-valued spectral measures associated to a root matrix.
This extension yields criteria for lower bounds on the multiplicity of absolutely continuous spectrum.
As applications, we consider random decaying shell-matrix potentials on stair-like graphs with increasing-width wires and operators on the Bethe lattice. 
In both cases, we establish that absolutely continuous spectrum has almost surely infinite multiplicity.


\end{abstract}

\maketitle

\tableofcontents

\section{Introduction}
Transfer matrices are among the fundamental tools in the spectral theory of one-dimensional and quasi-one-dimensional operators.
In the classical one-dimensional setting, spectral averaging formulas for Jacobi operators, developed by Carmona and Lacroix \cite{CL}, provide a criterion for the existence of absolutely continuous spectrum (in the following sometimes abbreviated by a.c. spectrum) and have played an important role in the study of random operators  \cite{CL, KP, KLS, LS}.
For ergodic models, the asymptotic growth of transfer matrix products is described by  
Lyapunov exponents whose positivity is a central ingredient 
 in many proofs of Anderson localization for one- and quasi-one-dimensional Anderson models \cite{GMP, KuSo,KLS,JitoZ,Bucaj-Fillman}. 
On the delocalization side, Kotani Theory \cite{K} relates the multiplicity of the absolutely continuous spectrum to the number of vanishing Lyapunov exponents \cite{KoS}.
Another important approach in the one-dimensional setting is subordinacy theory \cite{GP,KP}, which characterizes spectral types through the asymptotic behavior of transfer matrix products. Recent work \cite{WXZ} has extended this approach to long-range potentials in $\ell^2(\Z)$.
The global theory for one-frequency quasi-periodic Schr\"odinger operators developed by Avila \cite{Av} also stems from the analysis of the associated transfer matrix cocycle.

In higher dimensions, the situation is considerably less understood. A central open problem is the extended states conjecture, which asserts the existence of delocalized states, and in particular a.c. spectrum for the Anderson model on $\Z^d$, $d\geq 3$ at small disorder. This underlines the need for new methods in higher-dimensional spectral theory. 
The general physical picture, which may depend on the geometry and the connectivity of the underlying graph, predicts Anderson localization in dimensions $d=1,2$ at every disorder strength, in any dimension at sufficiently large disorder and near the edges of the spectrum. 
While localization in dimension one is rigorously established under very general assumptions on the single-site distribution
, localization for all disorder in dimension two has remained open.  
The principal methods for proving Anderson localization are the fractional moment method \cite{AM} and multi-scale analysis \cite{FS, GK1, GK2}. 
A particularly challenging case is the Anderson-Bernoulli model where localization for $d\ge2$ has presented substantial difficulties. 
Following the work of Bourgain and Kenig in the continuum setting \cite{BK}, localization has been proved in dimensions $d=2,3$ near the bottom of the spectrum and, partially, at large disorder \cite{DS,LZ, Li} and recently near the bottom in any dimension \cite{LLZ}.

In contrast,  the existence of continuous spectrum at sufficiently small disorder for Anderson models on $\ell^2(\Z^d)$ or $L^2(\R^d)$ remains an open problem. Delocalization has nevertheless been established for infinite-dimensional graphs of hyperbolic type, such as regular trees and tree-like structures \cite{Kl, ASW, AW, FHS, FHS2, KS, KLW, Sa-FC, Sa-Fib}. Recently, OpenAI released two manuscripts proposing proofs of Anderson localization in dimension $d=2$ at every positive disorder strength and Anderson delocalization in dimensions $d\geq 3$ at sufficiently small disorder, for independent uniformly distributed site potentials. These works are available in the OpenAI mathematics GitHub repository. At present, the correctness of these papers has not been established.

A significant recent work of Sadel \cite{Sa-Tr} introduces a framework of radial transfer matrix sets and generalized averaging formulas and 
criteria for a.c. spectrum for discrete Schrödinger operators on a locally finite countable graph \cite{Sa-OC,Sa-Tr,Sa-AT}. 
This extension is particularly remarkable since it makes transfer matrix techniques applicable to higher-dimensional graphs whereas such methods had traditionally been applied to one-dimensional and quasi-one-dimensional operators. 
The key idea is to replace the usual matrices by suitable sets of transfer matrices that encode the radial structure of the underlying graph.
The averaging formula was first established for one-channel operators \cite{Sa-AT, Sa-OC}, and more generally for Hermitian operators with locally finite hopping \cite{Sa-Tr}.
For unitary operators, the one-channel case has also been treated in \cite{BMST} and the multi-channel case is the subject of the ongoing work \cite{BST}.
These developments highlight the potential of transfer matrix methods beyond their traditional setting and thus motivate further extensions. In particular, it is natural to investigate whether the spectral averaging approach can provide information not only about the existence of a.c. spectrum, but also about its multiplicity. 

The purpose of the present work is to extend this framework from scalar to matrix-valued spectral measures associated with a root
matrix. One may hope to eventually apply this to the multi-dimensional Anderson model.
From a physical point of view, delocalized random operators in higher dimensions should conduct in all directions like the Laplacian and the absolutely continuous spectrum is expected to have infinite multiplicity.

The analysis in \cite{Sa-Tr} is based on a single root vector   $\Upsilon_0\in \Hi=\ell^2(\G)$, at which the spectral measure is computed.
It is therefore restricted to the cyclic subspace generated by this vector and does not give any information about the multiplicity of the spectrum.
To overcome this limitation, we consider a finite collection of root vectors  arranged as the columns of a root matrix $\Upsilon_0\in\Hi^{r_0}\cong\Ll(\C^{r_0}\to\Hi)$ and study the associated 
$r_0\times r_0$ matrix-valued spectral measure defined by
$$
\m(f)\,=\,\Upsilon_0^* f(H) \Upsilon_0,
$$
for bounded Borel functions $f$ where $H$ is a Hermitian operator with locally finite hopping.
Our main result is a matrix-valued version of Sadel's spectral averaging formula. In addition, we derive a criterion for a.c. spectrum of full multiplicity $r_0$ on the cyclic subspace generated by the columns of 
$\Upsilon_0$ and hence lower bounds on the multiplicity of the a.c. spectrum of $H$.
We apply this criterion to Anderson-type models on  stair-like graphs with an increasing number of wires and the Bethe lattice, in both cases with
random decaying shell-matrix potentials.
In these models, we show that the absolutely continuous spectrum has almost surely infinite multiplicity.

Beyond these applications, we expect the criterion to be useful for other random operators and for generalizations of Kotani Theory based on pseudo-Lyapunov exponents defined through the dynamics of the radial transfer matrix sets.

\subsection{Some conventions and notation}
Before recalling the set-up of \cite{Sa-Tr}, let us introduce some notation and identifications we use throughout the paper.
We utilize a double-stroke notation for (important) sets.
In particular, $\G$ will denote some countable set (or graph) and we will consider the Hilbert space $\Hi=\ell^2(\G)$
with the standard physics inner product 
$$
\langle \Psi,\Phi\rangle\,=\,\sum_{x\in\G} \overline{\Psi(x)} \Phi(x)\,.
$$

\vspace{.2cm}

With $\A^\B$ we denote the set of functions from $\B$ to $\A$. Elements of $\C^\G$  may be regarded as generalized states and we have $\Hi\subset \C^\G$. Furthermore,  for finite subsets $\Sb\subset \G$ we have $\ell^2(\Sb)=\C^\Sb$ which can be identified with $\C^{|\Sb|}$ and its standard inner product $\langle v,w\rangle=v^*w$ by choosing any orthonormal basis of $\C^\Sb$.

\vspace{.2cm}

Whenever  $\A\subset \B$, an element $f\in \C^\A$ will be identified with an element $f\in\C^\B$ by extending the function with the value 0, i.e. $f(b)=0$ for $b\in\B\setminus\A$.
This way, for $\Sb\subset \G$, the space $\ell^2(\Sb)$ will be considered as a subspace of $\Hi$ identifying $\psi\in\ell^2(\Sb)$ with $\psi\oplus\nul\in\Hi$.
Thus, if $\A\subset\Sb\subset \G$ with $|\Sb|<\infty$, we identify as subsets,  $\C^\A\subset\C^\Sb\subset\Hi$, to avoid the use of clumsy notation with natural projections and injections.

\vspace{.2cm}

Given two finite sets $\A,\B$, we identify $\C^{\A\times\B}$ with the set of linear maps from $\C^\B$ to $\C^\A$ in the natural way,
$(V(\chi))(a)=\sum_{b\in\B} V(a,b) \chi(b)$ for $V\in\C^{\A\times \B}$. 
We may also interpret $V$ as a collection of $\C^\A$-(column-)vectors $(v_b)_{b\in\B}$, with $v_b(a)=V(a,b)$, then
$V(\chi)=\sum_b \chi(b) v_b$.
Note that with the identifications of subsets as above (extensions with value 0), for $\tilde \A\subset \A\subset \G$ and $ \tilde \B\subset\B\subset \G$, $|\A|<\infty$, $|\B|<\infty$ we have
$$
\C^{\tilde \A\times \tilde \B} \subset \C^{\A\times \B} \subset \Ll(\C^\B \to \Hi) \equiv \Hi^\B \qtx{and}   \C^{\A\times \B}\subset \Ll(\Hi\to\C^\A).
$$
Here, $\Ll(\Hi_1 \to \Hi_2)$ denotes the set of (bounded) linear maps from $\Hi_1$ to $\Hi_2$. 
\newline
Similarly, for $|\A|<\infty$ we use the notations $\C^{\A\times r}=\Ll(\C^r\to\C^\A)$ and $\C^{r\times \A}=\Ll(\C^\A\to\C^r)$. Furthermore, we identify $V\in \C^{\A\times r}$ with a collection of $r$ `column' vectors (or `ket' vectors) in $\C^\A$, $V=(v_1,\ldots,v_r)$, $v_j\in\C^\A$. Similarly, $V\in\C^{r\times \A}$ may be identified with a collection of $r$ `row' vectors (or $r$ `bra' vectors) in the dual of $\C^\A$.
Then, with these identifications as subsets, 
$$\C^{\tilde \A \times r}\subset \C^{\A \times r}\subset \Hi^r\equiv \Ll(\C^r\to\Hi),\quad \C^{r\times \tilde\A} \subset \C^{r\times \A} \subset \Ll(\Hi\to\C^r)\,.$$
These identifications fit perfectly with adjoints. For example, for $\Upsilon\in\C^{\A\times r}$ the adjoint $\Upsilon^*\in \C^{r\times \A}$ interpreted as an element in $\Ll(\Hi\to\C^r)$ corresponds to the adjoint of $\Upsilon$ interpreted as an element of $\Ll(\C^r\to\Hi)$.
Then, for $H\in\Ll(\Hi)$ the expression $\Upsilon^* H \Upsilon$ is a linear map from $\C^r$ to $\C^r$ which will be interpreted as an $r\times r$ matrix.

\subsection{The basic setup}
Following \cite{Sa-Tr}, let $H$ be a locally finite symmetric (possibly unbounded) operator on $\Hi=\ell^2(\G)$ defined on the set $\Dd_{cs}\subset \Hi$ of compactly (i.e. finitely) supported functions from $\G$ to $\C$ and let us assume that $H$ is essentially self-adjoint on $\Dd_{cs}$.
Locally finite hopping means that for any $x\in\G$ the set 
$$
\Nn_x:=\{y\in\G\,:\,\langle\delta_y, H\delta_x\rangle\neq 0\}
$$
is finite. As $H$ has locally finite hopping, it naturally extends to an operator on $\C^\G$ and since it is essentially self-adjoint on $\Dd_{cs}$ then it is self-adjoint on its maximal domain in $\Hi=\ell^2(\G)$ given by
$$
\Dd=\{\Psi\in\Hi\,:\,H\Psi\in\Hi\}\,.
$$ 

As in \cite{Sa-Tr} we may associate a graph structure with an edge between $x$ and $y\neq x$ iff $\langle \delta_y,H\delta_x\rangle\neq 0$.
With this graph structure, we assume that $\G$ is connected.
Then, one may construct a quasi-spherical partition of $\G$ into finite sets $\Sb_n$, $n\in\Z_+=[0,\infty)\cap\Z$
such that $\Sb_n$ only connects to $\Sb_n$ and $\Sb_{n\pm1}$.
In other words, for
$x\in \Sb_j,\, y\in\Sb_k$ and $|k-j|>1$ we have $\langle\delta_y, H \delta_x\rangle=0$.
For instance, choosing any finite non-empty set $\Sb_0$ and letting $\Sb_n=\{x:d(x,\Sb_0)=n\}$ where $d(x,\Sb_0)$ is the graph distance of $x$ from $\Sb_0$, the resulting partition satisfies this condition.

According to this partition one has an orthogonal sum structure of the Hilbert space,
$$
\G=\bigsqcup_{n=0}^\infty \Sb_n,\quad \Hi=\ell^2(\G)=\bigoplus_{n=0}^\infty \C^{\Sb_n},
$$
and thus the operator $H$ is tri-diagonal in blocks
\begin{equation} \label{eq-H0}
H\,=\,\pmat{V_0 & W_1^* \\ W_1 & V_1 & W_2^*\\ & W_2 & V_2 & \ddots \\ & & \ddots & \ddots},
\end{equation}
where $V_n \in \Her(\C^{\Sb_n})$ is a Hermitian matrix in $\C^{\Sb_n\times \Sb_n}$ and
$W_n\in \C^{\Sb_n\times \Sb_{n-1}}$.
We let $r_n=\rank(W_n)$ and use a singular value type decomposition given by
$$
W_n\,=\,-\Upsilon_n \Phi_{n-1}^*\qtx{where} \Upsilon_n\in\C^{\Sb_n\times r_n}, \Phi_{n-1} \in \C^{\Sb_{n-1}\times r_n} \qtx{for} n\geq 1\,.
$$
$\Upsilon_n$ and $\Phi_{n-1}$ necessarily have full rank $r_n$. 
Using the orthogonal sum structure and letting $\psi=\bigoplus_{n=0}^\infty \psi_n$ with $\psi_n\in\C^{\Sb_n}$
and the convention $\Phi_{-1}^*\psi_{-1}=0$ we have
\begin{equation}\label{eq-H}
(H\psi)_n\,=\,-\,\Phi_n \Upsilon_{n+1}^* \,\psi_{n+1}\,-\,\Upsilon_{n}\,\Phi_{n-1}^*\psi_{n-1}\,+\,V_n\,\psi_n\;.
\end{equation}
Interpreting $\Upsilon_n$, $\Phi_{n-1}$ as a collection of column vectors, $\Upsilon_n=(\Upsilon_{n,1},\ldots,\Upsilon_{n,r_n})$, $\Phi_{n-1}=(\Phi_{n-1,1},\ldots,\Phi_{n-1,r_n})$ we can write $H$ in Dirac notation as follows:
$$
H\,=\, - \sum_{n=1}^\infty \left(\sum_{j=1}^{r_n} |\Upsilon_{n,j}\rangle \langle \Phi_{n-1,j}| + | \Phi_{n-1,j}\rangle \langle\Upsilon_{n,j}| \right)\,+\,\bigoplus_{n=0}^\infty V_n\,.
$$
Note that there is actually no $\Upsilon_0$ appearing in the operator $H$ and we are completely free to choose it.
Thus we let $\Upsilon_0\in\C^{\Sb_0\times r_0}$ be of rank $r_0$ where $r_0\leq r_1$.
We will study the $r_0\times r_0$ matrix-valued spectral measure at $\Upsilon_0$ defined by
\begin{equation}\label{eq-mu}
\m(f)\,:=\,\Upsilon_0^* f(H) \Upsilon_0\,.
\end{equation}

\subsection{Transfer matrices}
For $m\leq n$ consider the subsets 
$$
\G_{m,n}\,:=\,\bigsqcup_{k=m}^n \Sb_k \subset \G
$$
and by $H_{m,n}\in\Her(\C^{\G_{m,n}})$ we denote the Hermitian matrix obtained from the restriction of $H$ to $\C^{\G_{m,n}}$. Using \eqref{eq-H0}, this means $H_{m,n}$ corresponds to the tri-diagonal block-matrix from $V_m$ to $V_n$ on its diagonal.
Physicists may view this as using Dirichlet-type boundary conditions.
Then, we define the $m$ to $n$ boundary data, $m\leq n$, at the spectral parameter $z \not \in \spec(H_{m,n})$ by
\begin{equation}
R^z_{m,n}=\pmat{\alpha^z_{m,n} & \beta^z_{m,n} \\ \gamma^z_{m,n} & \delta^z_{m,n}}\,:=\,\pmat{  \Upsilon_m^* \\ \Phi_n^*} (H_{m,n}-z)^{-1}\,\pmat{ \Upsilon_m & \Phi_n}
\end{equation} 
where
$$
\alpha_{m,n}^z\,\in\,\C^{r_m \times r_m}\;,\quad \beta^z_{m,n}\,\in\,\C^{r_m \times r_{n+1}}\;,\quad
\gamma^z_{m,n}\,\in\,\C^{r_{n+1} \times r_m}\;,\quad \delta^z_{m,n}\,\in\,\C^{r_{n+1}\times r_{n+1}}\;.
$$
This means, e.g., that $\alpha^z_{m,n}=\Upsilon^*_m  (H_{m,n}-z)^{-1}  \Upsilon_m$. 
Also note the general relations
\begin{equation}
\alpha^{\bar z}_{m,n}\,=\,(\alpha^z_{m,n})^*\;, \quad
\gamma^{\bar z}_{m,n} = (\beta^z_{m,n})^*\;,\quad \delta^{\bar z}_{m,n}=(\delta^z_{m,n})^*\;.
\end{equation}
For $m=n$ we shorten the notations to
\begin{equation}
R^z_{n}=\pmat{\alpha^z_n & \beta^z_n \\ \gamma^z_n & \delta^z_n}\,:=\,\pmat{\Upsilon_n^* \\ \Phi_n^*}\,(V_n-z)^{-1}\,\pmat{\Upsilon_n & \Phi_n}\,.
\end{equation}
For some $z\in\spec(H_{m,n})$ one may define $R^z_{m,n}$ by analytic extension of $z\to R^z_{m,n}$.
The set of $z$ where $R^z_{m,n}$ can not be defined is denoted by
\begin{equation}
\A^1_{m,n}\,:=\,\{\lam\in \R\,:\,R^\lam_{m,n}\;\text{is not defined after analytic extension}\;\}\;,\quad \A^1_n\,:=\,\A^1_{n,n}\;.
\end{equation}
Clearly, $\A^1_{m,n}$ is a subset of the eigenvalues of $H_{m,n}$ and therefore finite.
As in \cite{Sa-Tr} we need the following assumptions:

\vspace{.2cm}

\noindent {\bf Assumptions:}\\
\noindent {\bf (A1):} $(r_n)_n$ is non-decreasing,
$
r_{n+1}\,\geq\;r_n\;.
$
\\
{\bf (A2):} We assume that for all $n\in\Z_+$ and almost all real $z=\lam \in \R$
 we have that the $r_n \times r_{n+1}$ matrix $\beta^\lam_n$ has full rank $r_n$.

\vspace{.2cm}  
  
The condition (A2) in principle says that the matrix $V_n$ connects all modes of $\Upsilon_n$ and $\Phi_n$. This will be the case for typical operators like Schrödinger operators (adjacency+potential) on a connected graph.
In fact, if $\beta^z_n$ has full rank for some real $z=\lam$, then by considering the rational function $z\mapsto \det( \beta^z_n (\beta^{\bar z}_n)^*)$ we see that it has full rank for all but finitely many $z\in \C$.
The condition (A1) can always be obtained after grouping shells together.
If the $\liminf$ of $r_n$ is bounded one can group shells together such that the rank is (eventually) constant\footnote{One has a subsequence $n_k$ with 
$r_{n_k}=r=\liminf r_n$ and one can define new shells $S'_k:=\G_{n_k,n_{k+1}-1}$ (defining $n_0=0$) to get a partition of $\G$ with ranks $r'_k=r$.}. 
This happens for strip-like structures and certain anti-trees. Otherwise, one can group the shells such that $r_n$ is increasing. Now, if one groups together $\Sb_m$ up to $\Sb_n$ (with $r_{n+1}\geq r_m$), then one needs
$r_i \geq r_m$ for any $m\leq i \leq n$ in order to have a chance of fulfilling (A2). But the grouping can always be done in such a way.

Under assumptions (A1) and (A2) we define
\begin{align}
 \A^2_{m,n} \,&:=\,\{z\,\in\,\C\setminus\A^1_{m,n}\,:\,\rank\,\beta^z_{m,n} < r_m\,\}\;,\\
 \A_{m,n} \,&:=\,\A^1_{m,n}\cup \A^2_{m,n}\;,\; \A_n\,:=\,\A_{n,n}\;.
\end{align}
With the assumptions above we immediately get that all $\A_n$ are finite sets. Furthermore, in \cite{Sa-Tr} it was shown that under these conditions, 
any set $\A_{m,n}$ is also finite. We call $\A_n$ the set of singular parameters at the shell $n$.

As in \cite{Sa-Tr} we define the sets of transfer matrices associated to $\G_{m,n}$ for $z\not \in A_{m,n}$ by
\begin{align}
\T^z_{n\ot m}\,&:=\,\left\{\pmat{B & - \bbb   \\ \delta^z_{m,n} B  & \gamma^z_{m,n}-\delta^z_{m,n} \bbb }\,:\, \beta^z_{m,n} B\,=\,\one_{r_n},\;
\beta^z_{m,n} \bbb\,=\,\alpha^z_{m,n} \right\}\;,  \label{eq-tr-1} \\
\;\;\T^z_n\,&:=\,\T^z_{n\ot n}\;. \label{eq-tr-1a}
\end{align}
Furthermore, for $z\not\in \A_{0,n}$  we define the set of boundary matrices at $n$ with initial Dirichlet condition
\begin{equation}
\D^z_{n}\,:=\,\left\{\pmat{B \\ \delta^z_{0,n} B}\,:\,\beta^z_{0,n} B\,=\,\one_{r_0} \right\}\,=\, \T^z_{0,n} \pmat{\one_{r_0} \\ 0} \,\subset\,\C^{2r_{n+1} \times r_0}\;
\end{equation}
as well as the set of boundary matrices with initial Neumann condition
\begin{equation}
\N^z_{n}\,:=\,\left\{\pmat{-\bbb \\ \gamma^z_{0,n}-\delta^z_{0,n} \bbb}\,:\,\beta^z_{0,n} \bbb\,=\,\alpha^z_{0,n} \right\}\,=\,  
\T^z_{0,n} \pmat{0 \\ \one_{r_0}} \,\subset\,\C^{2r_{n+1} \times r_0}\;.
\end{equation}
In these definitions, $\alpha^z_{m,n}, \beta^z_{m,n}, \gamma^z_{m,n}, \delta^z_{m,n}$ are fixed and we take the sets over all $B$ and $\bbb$ satisfying the conditions.

Note that $\T^z_{n\ot m}, \D^z_n, \N^z_n$ are affine subspaces.
As a matter of fact, if we define the subspaces
\begin{equation}
\VV^z_{m,n}=\left\{ \pmat{v\\ \delta^z_{m,n} v}\,:\,v\in\ker(\beta^z_{m,n}) \right\}\,\subset\, \C^{2r_{n+1}}
\end{equation}
and the induced matrix subspaces
\begin{equation}
\left(\VV^z_{m,n}\right)^{\oplus q}\,=\,\left\{(v_1,\ldots,v_q)\,:\,v_j\in\VV^z_{m,n}\right\}\,\subset\,\C^{r_{n+1}\times q}\,,
\end{equation}
then, for any $T\in\T^z_{n\leftarrow m},\, D\in \D^z_{n},\, N\in\N^z_{n}$ we have
\begin{equation}
\T^z_{n\leftarrow m}\,=\,T+(\VV^z_{m,n})^{\oplus 2r_m }\,,\quad
\D^z_n=D+(\VV^z_{0,n})^{\oplus r_0}\,,\quad
\N^z_n=N+(\VV^z_{0,n})^{\oplus r_0} \label{eq-str-T}\,.
\end{equation}

\begin{rem*} \label{rem-T}
Consider $\G=\Z_+\times \Sb$ with $|\Sb|$ finite
and let $\Sb_n=\{n\}\times \Sb$ , $\psi_n\in \C^{\Sb_n}\equiv\C^\Sb$ and consider an operator $H$ of the form
$$
(H\psi)_n\,=\,-\psi_{n-1}-\psi_{n+1}+V_n \psi_n\,,\quad n\geq 0,\quad \psi_{-1}=0\,.
$$
Then, for $n\geq 0$ we can select $\Upsilon_n=\Phi_n=\one$ giving
$\alpha^z_n=\beta^z_n=\gamma^z_n=\delta^z_n= (V_n-z\one)^{-1}$ and 
$$
\T^z_n=\{T^z_n\}\,,\quad T^z_n=\pmat{(\beta^z_n)^{-1} & -(\beta^z_n)^{-1} \alpha^z_n \\ \delta^z_n (\beta^z_n)^{-1}& \gamma^z_n-\delta^z_n(\beta^z_n)^{-1} \alpha^z_n}\,=\,\pmat{V_n-z\one & -\one \\ \one & \nul}
$$
which is the standard $n$-th transfer matrix for such a block Jacobi operator.
\end{rem*}

Before summarizing some basic properties of the transfer matrices, let us make some definitions.

\begin{definition}
Let $J_r=\pmat{\nul & -\one_r \\ \one_r & \nul}$ denote the $2r\times 2r$ symplectic form and let $q\geq r$ be two natural numbers and
$\T\subset \C^{2q\times 2r}$. We say that $\T$ is a {\bf Hermitian symplectic set} of matrices, if
for all $T_1, T_2 \in \T$ we have
$$
T_1^* J_q T_2\,=\,J_r\,.
$$
Similarly, we say that $\T$ is a {\bf symplectic set} of matrices, if for all $T_1, T_2 \in \T$
$$
T_1^\top J_q T_2\,=\,J_r\,.
$$
\end{definition}

For $\ell\leq m < n$ let
\begin{equation}
\B_{\ell,m,n}\,:=\,\left\{\lam \in \R\,:\, \one_{r_{m+1}}\,-\,\alpha^\lam_{m+1,n} \delta^\lam_{\ell,m}\;\;\text{is not invertible}\;\right\}
\end{equation}
which is a finite set as mentioned in \cite{Sa-Tr}.

For subsets of matrices $\T_1, \T_2$ we denote $\T_2 \T_1:=\{T_2T_1\,:\,T_i\in\T_i\}$ if the matrices can be multiplied this way, and
$\T_1+\T_2=\{T_1+T_2\,:\,T_i\in\T_i\}$ if the matrices can be added.
Let us summarize the properties proved in \cite{Sa-Tr}.

\begin{proposition}\label{prop-1}
Under assumptions {\rm (A1)} and {\rm (A2)} we have the following:
\begin{enumerate}[{\rm (i)}]
\item Let $\ell,m,n\in\Z_+$, $\ell\leq m <n$. Then, $\A_{\ell,n}$ is finite, $\A_{\ell,n}\subset \A_{\ell,m} \cup \A_{m+1,n} \cup \B_{\ell,m,n}$
and for $z\not \in \A_{\ell,m} \cup \A_{m+1,n} \cup \B_{\ell,m,n}$
$$
\T^z_{n \ot m+1} \T^z_{m \ot \ell} \pmat{\one_{r_\ell} \\ \nul}\,=\,
\T^z_{n \ot \ell} \pmat{\one_{r_\ell} \\ \nul}\,,\quad
\T^z_{n \ot m+1} \T^z_{m \ot \ell} \pmat{\nul \\ \one_{r_\ell}}\,=\,\T^z_{n \ot \ell} \pmat{\nul \\ \one_{r_\ell}}\,.
$$
In particular,
$$
\T^z_{n\ot m+1}\,\T^z_{m \ot \ell}\,\subset\, \T^z_{n \ot \ell}
$$
and for $z\not \in \A_{0,m} \cup \A_{m+1,n} \cup \B_{0,m,n}$ we have
$$
\T^z_{n \ot m+1} \D^z_{m}\,=\,\D^z_n\qtx{and}
\T^z_{n\ot m+1} \N^z_m\,=\,\N^z_n\,.
$$
This implies for fixed $n\in \Z_+$ and all but finitely many $z\in \C$ that
$$
\D^z_n\,=\,\T^z_{n\leftarrow 0}\pmat{\one_{r_0}\\ \nul}\,=\,\T^z_n \T^z_{n-1}\cdots \T^z_1 \T^z_0 \pmat{\one_{r_0}\\ \nul}\,,
$$
$$
\N^z_n\,=\,\T^z_{n\leftarrow 0}\pmat{\nul \\ \one_{r_0}}\,=\,\T^z_n \T^z_{n-1}\cdots \T^z_1 \T^z_0 \pmat{\nul \\ \one_{r_0}}\,.
$$
\item For $z=\lam \in \R \setminus \A_{m,n}$ the transfer matrix set
$\T^\lam_{n\ot m}$ is a Hermitian symplectic set, meaning, for any $T_1, T_2 \in \T^\lam_{n\leftarrow m}$ we have 
$T_1^* J_{r_{n+1}} T_2\,=\, J_{r_m}\,.$\\
If all $V_n$ and all $\Upsilon_n$ and $\Phi_n$ for $n\in\Z_+$ are real matrices, then for any $z\in\C \setminus \A_{m,n}$ the transfer matrix set $\T^z_{n \ot m}$ is a symplectic set, meaning for any $T_1, T_2 \in \T^z_{n\ot m}$
we find $T_1^\top J_{r_{n+1}} T_2= J_{r_m}\,.$

\item Let $z, \bar z \not \in \bigcup_{n=0}^\infty \A_n$  and let $\Psi=(\Psi_n)_{n=0}^\infty \in \C^\G$ be a formal solution of 
$H\Psi=z\Psi+ \Upsilon_0 v$, i.e.
$(H\Psi)_n=z \Psi_n$ for $n\geq 1$ and $(H\Psi)_0=z\Psi_0+ \Upsilon_0 v$ with $v \in \C^{r_0}$. 
Assume that $v\neq 0$ or $u:=\Upsilon_0^* \Psi_0 \neq 0$.
Then, there exist $T^z_n \in \T^z_n$, $n\in \Z_+$, such that
$$
\pmat{\Upsilon_{1}^* \Psi_{1} \\ \Phi_0^* \Psi_0 }\,=\, T^z_0 \pmat{u \\ v }\qtx{and}
\pmat{\Upsilon_{n+1}^* \Psi_{n+1} \\ \Phi_n^* \Psi_n}\,=\,T^z_n \,\pmat{\Upsilon_{n}^* \Psi_{n} \\ \Phi_{n-1}^* \Psi_{n-1}}\,.
$$
In particular, we find
$$
\pmat{\Upsilon_{n+1}^* \Psi_{n+1} \\ \Phi_n^* \Psi_n }\,=\, T^z_n T^z_{n-1} \cdots T^z_0\, \pmat{\Upsilon_0^* \Psi_0 \\ v }\,.
$$
\item Let  $z\not \in \bigcup_{n=0}^\infty \A_n $. Given a vector $\smat{u\\v}\in \C^{2r_0}$ and a selection of transfer matrices $T^z_n \in \T^z_n$, there exists a formal solution $\Psi=(\Psi_n)_{n=0}^\infty \in \prod_{n=0}^\infty \C^{S_n}=\C^\G$ of
$H\Psi=z\Psi+\Upsilon_0 v$ such that
$$
\Upsilon_0^* \Psi_0\,=\,u \qtx{and} \pmat{\Upsilon_{n+1}^* \Psi_{n+1} \\ \Phi_n^* \Psi_n }\,=\, T^z_n T^z_{n-1} \cdots T^z_0\, \pmat{u \\ v }\,.
$$

\end{enumerate}
\end{proposition}

Note that the correspondences in $(iii)$ and $(iv)$ are not one-to-one in general!
If $v=Wu=W \Upsilon_0^* \Psi_0$ for some Hermitian matrix $W$, then $\Psi$ is a formal solution of $(H-\Ups_0 W\Ups_0^* ) \Psi = z \Psi$
and one can relate to $W$ as some form of `boundary condition'.

Moreover, note that from \eqref{eq-str-T} it follows that in the sense of sums of sets,
\begin{equation}
\T^z_{n\leftarrow m}\,=\,\T^z_{n\leftarrow m} \pmat{\one_{r_m} & \nul \\ \nul & \nul}\,+\,\T^z_{n\leftarrow m} \pmat{\nul & \nul\\ \nul & \one_{r_m}}\,.
\end{equation}
Hence, by  Proposition~\ref{prop-1}~(i) it follows that
\begin{equation}
\T^z_{n \ot \ell}
\,=\,
\T^z_{n\ot m+1}\,\T^z_{m \ot \ell}\pmat{\one_{r_\ell}&\nul\\\nul&\nul}+\T^z_{n\ot m+1}\,\T^z_{m \ot \ell}\pmat{\nul&\nul\\\nul&\one_{r_\ell}}
\end{equation}
\begin{equation}
\T^z_{n\leftarrow 0}\,=\, \T^z_n \T^z_{n-1} \cdots  \T^z_0 \pmat{\one_{r_0} & \nul \\ \nul & \nul}\,+\,
 \T^z_n \T^z_{n-1} \cdots  \T^z_0 \pmat{\nul & \nul\\ \nul & \one_{r_0}}
\end{equation}

\subsection{Main Results}

Recall that we assumed $H$ to be essentially self-adjoint on $\Dd_{cs}$, the set of compactly supported functions.
This is equivalent to $H$ being self-adjoint on its maximal domain $\Dd$ as an operator on $\Hi$.
As mentioned above, we consider the matrix-spectral measure $\m$ at $\Upsilon_0$ given by
 $$
\m(f)\,=\,\int f(\lam)\,\m(\dd\lam)\,=\, \Upsilon_0^* f(H)\,\Upsilon_0 \;.
$$
As in \cite{Sa-Tr}, we need to separate the discrete part coming from compactly supported eigenfunctions. Therefore, we define the spaces
$$
\Hi_n\,=\,{\rm span}\,\left\{\psi\in \C^{\G_{0,n}}\subset \Hi\,:\,\psi\text{ is an eigenfunction of } H\right\}
$$
Note that $\Hi_n \subset \Hi_{n+1}$.
Then we let
$$
\Hi_{cs}=\overline{\bigcup_{n=0}^\infty \Hi_n}\,=\,\overline{{\rm span}\left\{ \psi\in\Dd_{cs}\,:\,\psi \text{ is an eigenfunction of } H \right\}}.
$$
Note that while $\Hi_{cs}$ is generated by compactly supported eigenfunctions of $H$, due to the closure, it may contain non-compactly supported functions on $\G$.
Moreover, the closed sub-spaces $\Hi_n$ and $\Hi_{cs}$ are left invariant by $H$, therefore, using the orthogonal projections,
we can define the corresponding parts of the measure $\m$.
Let $\Pp_n$, $\Pp_{cs}$ be the orthogonal projections of $\Hi$ onto $\Hi_n$ and $\Hi_{cs}$, respectively. Moreover, let $\Pp_{cs}^\perp$ be the orthogonal projection onto the orthogonal complement $\Hi_{cs}^\perp$.
Then define the matrix-valued measures
$$
\n_n(f)\,:=\,\Upsilon_0^* (\Pp_n f(H))\, \Upsilon_0\,=\,(\Pp_n\Upsilon_0)^*f(H) (\Pp_n \Upsilon_0)
$$
$$
\n(f)\,:=\,\Upsilon_0^*(\Pp_{cs} f(H))\, \Upsilon_0\,=\,(\Pp_{cs}\Upsilon_0)^*f(H) (\Pp_{cs} \Upsilon_0)
$$
$$
\n^\perp(f)\,:=\,\Upsilon_0^*(\Pp_{cs}^\perp f(H))\, \Upsilon_0\,=\,(\Pp_{cs}^\perp \Upsilon_0)^*f(H) (\Pp_{cs} \Upsilon_0)\,.
$$
Let us remark that by the definition, $\n_n\leq \n_{n+1}$ in the sense that for $f\geq 0$ we find that $\n_{n+1}(f)-\n_n(f)$ is positive semidefinite. Moreover, $\n_n \rightharpoonup \n$ weakly, and all $\n_n$ and $\n$ are point measures, and
$$
\m\,=\,\n\,+\,\n^\perp\,.
$$
Now, for
 $A\in \Her(\C^{r_{n+1}})$ define the following matrices and matrix-spectral measures at $\Upsilon_0$,
\begin{equation}
H_{0,n,A}\,=\,H_{0,n}\,-\, \Phi_n A \Phi_n^*\,,\qquad \m_{n}(A,f)\,=\,\Upsilon_0^*\, f(H_{0,n,A})\,\Upsilon_0
\end{equation}
and the average
\begin{equation}
\bar \m_n(f)\,=\,\int_{\R^{r_{n+1}}} \m_{n}(\diag(v_1,\ldots,v_{r_{n+1}}),f)\,\prod_j \frac{\dd v_j}{\pi(1+v_j^2)}
\end{equation}
Moreover, we let 
\begin{equation}\label{eq-def-nu_n^perp}
\n_n^\perp(f)\,=\,\bar\m_n(f)\,-\,\n_n(f)\,.
\end{equation}

Recall  that $\Dd_{cs}$ is a core for $H$.
As $H_{0,n,A_n}\psi \to H\psi$ for any $\psi\in\Dd_{cs}$, we have strong resolvent convergence of $H_{0,n,A_n}$ to $H$ as $n\to\infty$, implying
\begin{equation}\label{eq-limit}
\bar\m_n\rightharpoonup \m \qtx{and} \n^\perp_n \rightharpoonup \n^\perp \quad \text{weakly.}
\end{equation}
To present the results in a compact way, let us introduce some notations.
For closed matrix sets $\T\subset \C^{q\times r}$ we define
$$
\min \|\T\|^p\,:=\,\min\{ \|T\|^p\,:\,T\in\T\}\,
$$
using the standard matrix norm induced by the Euclidean vector norm.
By $|T|=\sqrt{T^*T}$ we denote the matrix absolute value of a matrix $T$.
By writing
$$
A=\max |\T|^{-2}=\left(\min |\T|^2\right)^{-1}
$$
we mean that
$$
\exists B\in\T\,:\,A=|B|^{-2}=(B^*B)^{-1} \qtx{and} \forall\, T\in\T\,:\,A\geq |T|^{-2}=(T^*T)^{-1}
$$
where the latter inequality is understood in the sense  that $A-|T|^{-2}$ is positive semidefinite.

First, let us mention an analogue of \cite[Theorem~4~(ii)]{Sa-Tr} including multiplicity of the spectrum.

\begin{theorem}\label{Theo}
Assume that $H$ is self-adjoint with its natural domain and that assumptions (A1), (A2) hold.
Suppose for some $p>1$, $[a,b]\subset \R$  one finds 
$$
\liminf_{n\to\infty} \int_a^b \min \| \T^\lam_{n\leftarrow 0}  \|^{2p}\,\dd\lam\,<\,\infty
$$
 then $[a,b] \subset \supp \m_{ac} \subset \spec_{ac}(H)$ and the matrix measure $\n^\perp=\m-\n$ is purely absolutely continuous in $(a,b)$ with an $L^p$ positive-semidefinite matrix density $\bff(\lam)$ which is invertible for Lebesgue almost all
 $\lam \in(a,b)$.
In particular,  within the cyclic space generated by the $r_0$ column vectors of $\Upsilon_0$, the multiplicity of the a.c. spectrum is exactly $r_0$ inside $[a,b]$.
Thus, the absolutely continuous spectrum of $H$ has at least multiplicity $r_0$ everywhere in $[a,b]$.
\end{theorem}
Note that $\lam\mapsto \min \| \T^\lam_{n\leftarrow 0}  \|$ is measurable which will be shown in the proof.
We also get the following analogue of \cite[Theorem~3]{Sa-Tr}.

\begin{theorem}\label{Theo-2}
Assume that $H$ is self-adjoint with its natural domain and that assumptions (A1), (A2) hold.
Let $a<b$ and $[a,b]\subset\R$ be given. Suppose that, for some increasing sequence $(n_k)_k$ of positive integers 
we find 
$$
\sup_{k\in\NN} \left( \min \left\| \D^\lam_{n_k} \right \| \right) \,<\,\infty\quad \text{for Lebesgue almost all $\lam\in(a,b)$}\;.
$$
Then, $[a,b]\subset \supp \m_{ac} \subset \spec_{ac}(H)$, and the matrix density $\bff(\lam)$ of its  absolutely continuous part is almost surely invertible on $(a,b)$.
In particular,  within the cyclic space generated by the $r_0$ column vectors of $\Upsilon_0$, the multiplicity of the a.c. spectrum is exactly $r_0$ inside $[a,b]$.
Thus, the absolutely continuous spectrum of $H$ has at least multiplicity $r_0$ everywhere in $[a,b]$.
\end{theorem}

As in the previous works \cite{Sa-AT, Sa-OC,Sa-Tr}, this is achieved by generalizing a spectral averaging formula of Carmona, using weak convergence arguments and the Banach-Alaoglu Theorem.
The spectral averaging formula in this case can be formulated in the following way.

\begin{theorem}\label{Theo-av} Assume that $H$ is self-adjoint with its natural domain and that assumptions (A1), (A2) hold.
Then,  $\n$ is supported on $\displaystyle\liminf_n \A^1_{0,n}=\bigcup_{n\geq 0} \bigcap_{m\geq n} \A^1_{0,m}$ and 
we find for bounded continuous functions $f\in C_b(\R)$ that
\begin{align*}
\n_n^\perp(f)\,&=\,\frac{1}{\pi} \int_\R f(\lam) \left| \left(1+(\delta^\lam_{0,n})^2\right)^{-1/2} \gamma^\lam_{0,n}\,\right|^2 \,\dd\lam \\
&=\, \frac1\pi \int_\R f(\lam) \max\,\left| \D^\lam_{0,n}  \right|^{-2}\;\dd\lam\\
&=\,\frac1\pi \int_\R f(\lam) \max \,\left| \T^\lam_n \T^\lam_{n-1}\cdots \T^\lam_{1} \T^\lam_0 \pmat{\one_{r_0}\\\nul} \right|^{-2}\;\dd\lam\,.
\end{align*}
Here, $\dd\lam$ denotes the Lebesgue measure in $\R$.
\end{theorem}

In particular, by \eqref{eq-limit} for $f\in C_b(\R)$
\begin{equation}
\m(f)\,=\,\n(f)\,+\,\lim_{n\to\infty} \frac1\pi \int_\R f(\lam) \max \,\left| \T^\lam_n \T^\lam_{n-1}\cdots \T^\lam_{1} \T^\lam_0 \pmat{\one_{r_0}\\\nul} \right|^{-2}\;\dd\lam\,.
\end{equation}
Note that this is a generalization of  \cite[Theorem~III.3.2]{CL}, \cite[Theorem~2.3(iii)]{Sa-AT} and \cite[Theorem~2(i)]{Sa-OC}.

The rest of  the paper is organized as follows. In Section~\ref{sec-App}, we introduce two discrete random models and state the results on their absolutely continuous spectral multiplicity, 
illustrating an application of Theorem~\ref{Theo}. Section~\ref{sec-av} establishes the spectral averaging formula (Theorem~\ref{Theo-av}), which underlies the criteria developed in Theorems~\ref{Theo} and~\ref{Theo-2}. Section~\ref{sec-mult} is devoted to the proofs of these theorems. Finally, Section~\ref{ApplicationRM} provides the proofs of the multiplicity results for the two models introduced in Section~\ref{sec-App}.

\section{Applications \label{sec-App}}

This section applies our criterion for the multiplicity of the a.c. spectrum to random models, extending the results in \cite[Section~2.2]{Sa-Tr}.

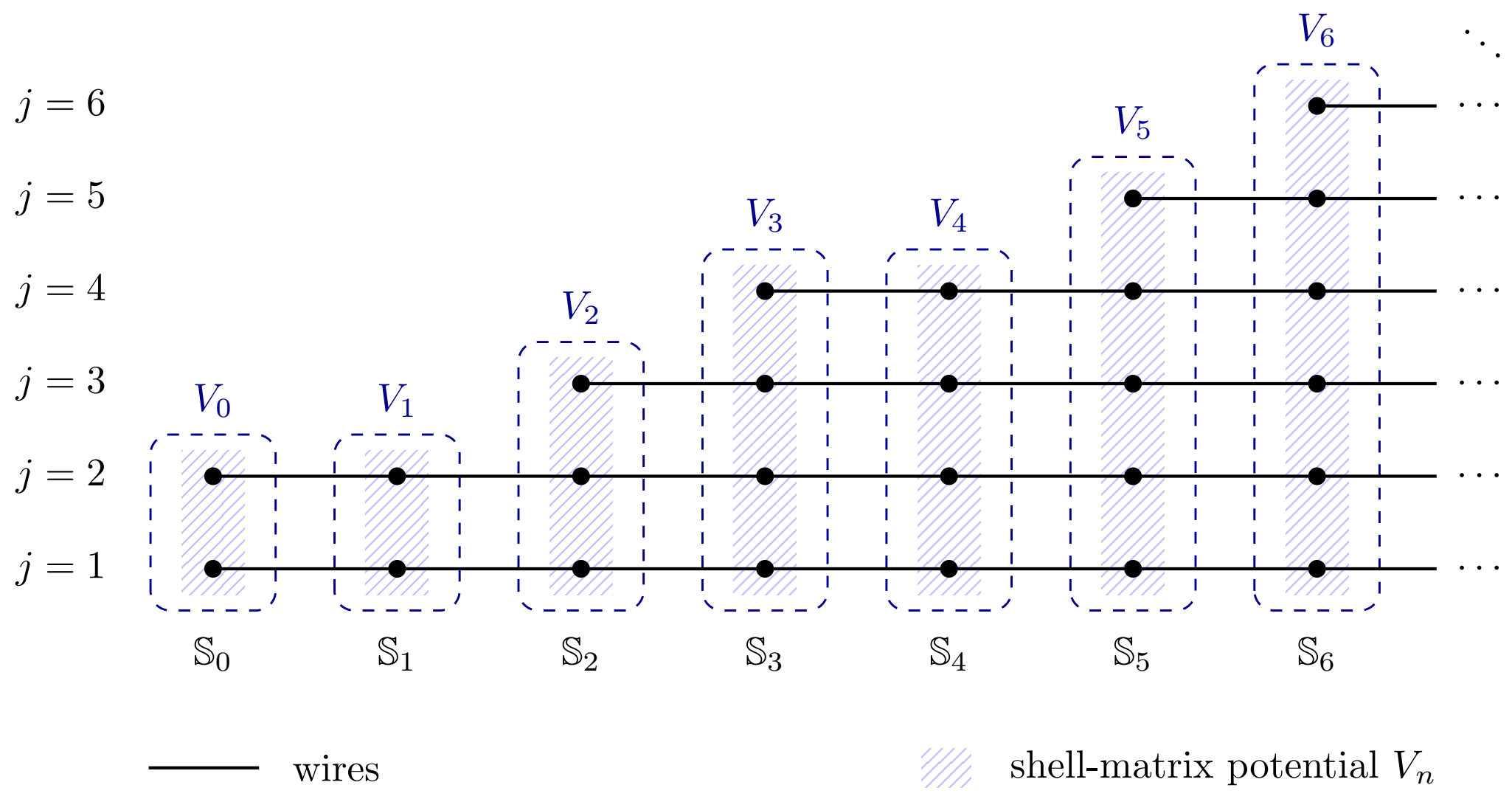
\begin{figure}[htbp]
\centering
\begin{tikzpicture}[
  x=1.65cm,y=0.83cm,
  vertex/.style={circle,fill=black,inner sep=1.6pt},
  wire/.style={black,line width=0.8pt},
  shell/.style={draw=blue!55!black,dashed,rounded corners=5pt,
                line width=0.55pt},
  potential/.style={fill=blue!8,pattern=north east lines,
                    pattern color=blue!25,draw=none},
  every node/.style={font=\small}
]
\foreach \n/\s in {0/2,1/2,2/3,3/4,4/4,5/5,6/6}{
  \path[potential] (\n-0.17,0.72) rectangle (\n+0.17,\s+0.28);
  \draw[shell] (\n-0.34,0.55) rectangle (\n+0.34,\s+0.45);
  \node[above,text=blue!55!black] at (\n,\s+0.48) {$V_{\n}$};
  \node at (\n,0.08) {$\Sb_{\n}$};
  \node[font=\scriptsize,text=black!65] at (\n,-0.38) {};
}

\foreach \j/\firstshell in {1/0,2/0,3/2,4/3,5/5,6/6}{
  \draw[wire] (\firstshell,\j) -- (6.65,\j);
  \node at (6.90,\j) {$\cdots$};
  \node[anchor=east] at (-0.50,\j) {$j=\j$};
}

\foreach \n/\s in {0/2,1/2,2/3,3/4,4/4,5/5,6/6}{
  \foreach \j in {1,...,\s}{
    \node[vertex] (v-\n-\j) at (\n,\j) {};
  }
}
\node[above right,font=\scriptsize] at (v-4-2) {};
\node at (6.90,6.8) {$\ddots$};

\draw[wire] (-0.35,-1.15) -- (0.25,-1.15);
\node[anchor=west] at (0.35,-1.15) {wires};
\path[potential] (3.85,-1.36) rectangle (4.12,-0.94);
\node[anchor=west] at (4.25,-1.15) {shell-matrix potential $V_n$};
\node[anchor=west] at (-0.35,-1.95)
  {};
\end{tikzpicture}
\caption{Independent wires are coupled by a random potential. The blue bars represent the shells $\Sb_n$ on which the potential $V_n$ acts.}
\label{fig:stair-like-random-model}
\end{figure}

Our first application considers a random decaying shell potential that couples an increasing number of wires. We begin by describing the model. Let $(s_n)_{n=0}^\infty$ be a non-decreasing sequence of natural numbers such that $s_n>0$, $\displaystyle\lim_{n\to\infty} s_n=\infty$ and define
$$
\G_1:=\lbrace\, (n,j)\,:\, n\in\N_0\,,\quad  j\in\lbrace 1,..., s_n\rbrace\rbrace\,\quad \Sb_n:=\lbrace n \rbrace\times\lbrace 1,...,s_n \rbrace\,.
$$
In the vector space $\ell^2(\G_1)$ we consider the free Hamiltonian given by
$$
\left(\Delta^{(1)}\psi\right)_{n,j}=-\psi_{n-1,j}- \psi_{n+1,j}+a_j\psi_{n,j}\,,\quad\text{for}\quad j\in\lbrace 1,...,s_n\rbrace\,,
$$
where $(a_j)_{j\in\N}$ is a sequence of real numbers satisfying $\sup_j a_j-\inf_j a_j<4$.
Moreover, for $(\psi_{n.j})_{(n,j)\in\G_1}\in\ell^2(\G_1)$ we define $\psi_{n-1,j}=0$ if $j>s_{n-1}$ or $n=0$. This operator describes independent wires at different mean energies $a_j$ for the $j$-th wire. More precisely, the $j$-th wire starts at the $n$-th shell where $s_{n-1}\leq j \leq s_n$. To couple these wires, we introduce a shell potential of the form
$$
V=\bigoplus_n V_n\quad\text{with}\quad V_n \in\Her(\C^{\Sb_n})\,.
$$
We further assume that $V_n=V_n(\omega)$ are independent $\Her(\C^{\Sb_n})$-valued random variables defined on some abstract probability space $(\Omega,\mathcal A, \mathbb P)$ satisfying 
\begin{equation}
\label{decaying-1}
\sum_{n=1}^\infty\left(\left\|\mathbb E(V_n)\right\|+\mathbb E\left(\|V_n\|^2+\|V_n\|^4\right)\right)<\infty\,,
\end{equation}
where we use the standard norm of $\C^{\Sb_n\times\Sb_n}$ and $\mathbb E$ denotes the expectation value. In this setting, we define the following random operator:
$$
\left(H^{(1)}_\omega\psi\right)_n:=\left(\Delta^{(1)}\psi\right)_n+V_n(\omega)\psi_n\,,\quad\text{where}\quad\psi_n=(\psi_{n,1},...,\psi_{n,s_n})^\top\in\C^{\Sb_n}\,.
$$
The resulting graph associated with this model is illustrated in Figure~\ref{fig:stair-like-random-model}.

Let $I_1\subset\R$ be the non-empty set given by 
$$
I_1:=\bigcap_{j=1}^\infty(-2+a_j, 2+a_j)\,.
$$
As a direct consequence of our criterion for the multiplicity, we have the following result:
\begin{theorem}\label{application_1}
Almost surely, the operator $H^{(1)}_{\omega}$ has purely absolutely continuous spectrum in $I_1$. Moreover, for almost all $\omega \in \Omega$ 
and $\lam\in I_1$,
we have
$$
m_{ac}\left(H^{(1)}_\omega,\lam\right)=\infty\,,
$$
where $m_{ac}(H,\lam)$ denotes the essential multiplicity of the a.c. spectrum of $H$ at energy $\lam \in \sigma_{ac}(H)$.
\end{theorem}

\begin{rem*}
Combining techniques of this paper with techniques from \cite{GS} one may omit the assumption $\sup_ja_j-\inf_j a_j<4$ and one may get absolutely continuous spectrum on the union 
$\bigcup_j (-2+a_j, 2+a_j)$. More precisely, given $N\subset \NN$ such that the set
$$
I_N\,=\,\bigcap_{j\in N} (-2+a_j, 2+a_j)
$$
is not empty, one can show that for almost all $\omega$, within $I_N$, the multiplicity of the absolutely continuous spectrum is at least $|N|$.
This will be dealt with elsewhere.
\end{rem*}

For our second application, we consider the Bethe lattice, here denoted by $\G_2$. This lattice is defined recursively: starting from a root vertex $0\in\G_2$ representing the zeroth generation, every vertex in the $n$-th generation is connected to two `children' in the $(n+1)$-st generation. 

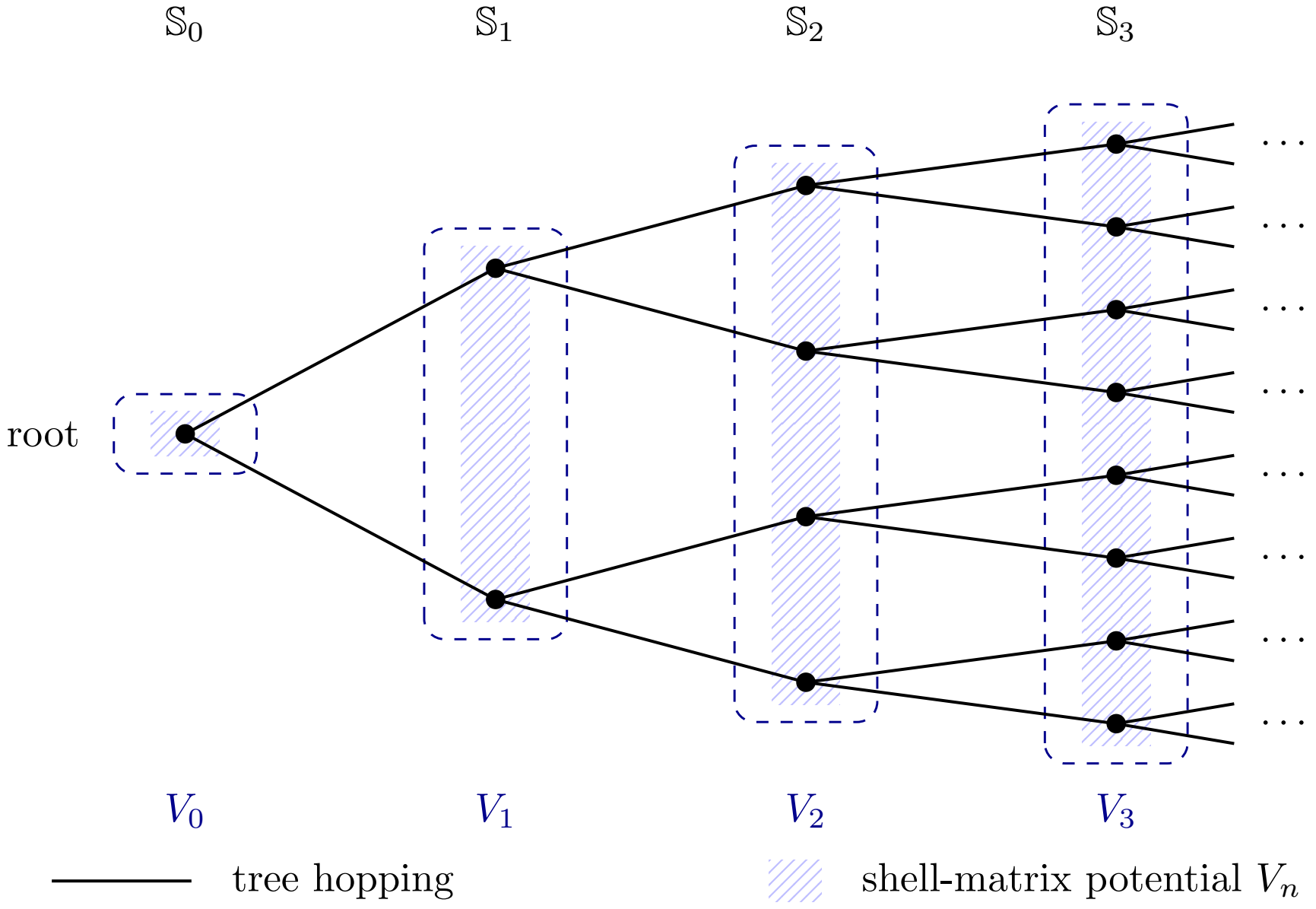
\begin{figure}[htbp]
\centering
\begin{tikzpicture}[
  x=2.7cm,y=0.72cm,
  vertex/.style={circle,fill=black,inner sep=1.7pt},
  edge/.style={black,line width=0.75pt},
  shell/.style={draw=blue!55!black,dashed,rounded corners=5pt,
                line width=0.55pt},
  potential/.style={fill=blue!8,pattern=north east lines,
                    pattern color=blue!25,draw=none},
  every node/.style={font=\small}
]
\foreach \n in {0,...,3}{
  \pgfmathtruncatemacro{\size}{2^\n}
  \pgfmathsetmacro{\height}{4-4/\size}
  \path[potential] (\n-0.11,-\height-0.27)
                    rectangle (\n+0.11,\height+0.27);
  \draw[shell] (\n-0.23,-\height-0.48)
                rectangle (\n+0.23,\height+0.48);
  \node at (\n,4.95) {$\Sb_{\n}$};
  \node[font=\scriptsize,text=black!65] at (\n,4.45)
    {};
  \node[text=blue!55!black] at (\n,-4.55) {$V_{\n}$};
  \foreach \j in {1,...,\size}{
    \pgfmathsetmacro{\yy}{4-(\j-0.5)*8/\size}
    \coordinate (v-\n-\j) at (\n,\yy);
  }
}

\foreach \n in {0,...,2}{
  \pgfmathtruncatemacro{\size}{2^\n}
  \pgfmathtruncatemacro{\nextlevel}{\n+1}
  \foreach \j in {1,...,\size}{
    \pgfmathtruncatemacro{\childa}{2*\j-1}
    \pgfmathtruncatemacro{\childb}{2*\j}
    \draw[edge] (v-\n-\j) -- (v-\nextlevel-\childa);
    \draw[edge] (v-\n-\j) -- (v-\nextlevel-\childb);
  }
}

\foreach \j in {1,...,8}{
  \pgfmathsetmacro{\yy}{4-(\j-0.5)}
  \draw[edge] (v-3-\j) -- (3.38,\yy+0.24);
  \draw[edge] (v-3-\j) -- (3.38,\yy-0.24);
  \node at (3.55,\yy) {$\cdots$};
}

\foreach \n in {0,...,3}{
  \pgfmathtruncatemacro{\size}{2^\n}
  \foreach \j in {1,...,\size}{
    \node[vertex] at (v-\n-\j) {};
  }
}
\node[anchor=east] at (-0.29,0) {root};
\node[above left,font=\scriptsize] at (v-1-1) {};
\node[below left,font=\scriptsize] at (v-1-2) {};

\draw[edge] (-0.43,-5.40) -- (0.02,-5.40);
\node[anchor=west] at (0.10,-5.40) {tree hopping};
\path[potential] (1.88,-5.65) rectangle (2.05,-5.15);
\node[anchor=west] at (2.13,-5.40) {shell-matrix potential $V_n$};
\node at (1.45,-6.25)
  {};
\end{tikzpicture}
\caption{The black edges represent the tree branches, and the blue strips indicate the layers $\Sb_n$ upon which the potential $V_n$ acts.}
\label{fig:bethe-lattice-random-model}
\end{figure}

The shell $\Sb_n$ is defined as the set of vertices in the $n$-th generation, meaning it consists of $2^n$ vertices.  The random potential is defined exactly as in the previous application, that is $V=\bigoplus_n V_n$ with $V_n(\omega)$ being independent $\Her(\C^{\Sb_n})$-valued random variables satisfying \eqref{decaying-1}. The free Hamiltonian is now taken to be 
$$
\left(\Delta^{(2)}\psi\right)_{n,j}:=-\psi_{n-1,\lceil j/2 \rceil}-\psi_{n+1,2j-1}-\psi_{n+1,2j}\,,
$$
where $\psi_{n-1,j}=0$ for $n=0$, and $\lceil j/2 \rceil$ denotes the smallest integer number greater than or equal to $j/2$. Consequently, the second random operator under consideration is given by
$$
\left( H^{(2)}_\omega\psi\right)_n=\left(\Delta^{(2)}\psi\right)_n+V_n(\omega)\psi_n\,,\quad \psi_n=(\psi_{n,1},...,\psi_{n,2^n})^\top\,.
$$
The graph associated with this model is shown in Figure~\ref{fig:bethe-lattice-random-model}.

Applying our multiplicity criterion, we obtain the following result:

\begin{theorem}\label{application_2}
Almost surely, the operator $H^{(2)}_{\omega}$ has purely absolutely continuous spectrum in the interval $I_2:=(-2\sqrt{2}\,,\, 2\sqrt{2})$, with $m_{ac}\left(H^{(2)}_\omega,\lam\right)=\infty$ for $\lam \in I_2$.
\end{theorem}

Theorems \ref{application_1} and \ref{application_2} generalize their counterparts in \cite[Section~2.2]{Sa-Tr}. As will be shown in Section~\ref{ApplicationRM}, using a root matrix $\Upsilon_0$ instead of a root vector provides a framework where the transfer matrix formalism extracts multiplicity information.

\section{Matrix-spectral averaging \label{sec-av}}

In this section we prove Theorem~\ref{Theo-av}. 
We start by analyzing the averaged spectral measure  $\bar\m_n$ through its matrix-valued Stieltjes transform $S_{\bar\m_n}(z)$, which is given by
$$
S_{\bar\m_n}(z)=   \int_{\R^{r_{n+1}}}\Upsilon_0^*\left(H_{0,n}- \Phi_n \diag(v_1,\ldots,v_{r_{n+1}}) \Phi_n^*-z \right)^{-1}\Upsilon_0\prod_j \frac{\dd v_j}{\pi(1+v_j^2)}\,.
$$
This function can be written in terms of the boundary data $R^z_{0,n}$. Indeed, using the fact that the Hermitian imaginary part of $\delta_{0,n}^z$ is positive definite together with the resolvent identity, one readily obtains (cf. \cite[Section~6]{Sa-Tr})
\begin{align}
& \Upsilon_0^*\left(H_{0,n}- \Phi_n \diag(v_1,\ldots,v_{r_{n+1}}) \Phi_n^*-z \right)^{-1}\Upsilon_0\,=\, \nonumber \\ 
&\qquad \qquad\qquad \qquad  \alpha_{0,n}^z+\beta_{0,n}^zA(v)(\one_{r_{n+1}}-\delta_{0,n}^zA(v))^{-1}\gamma_{0,n}^z\,=\,\\
&\qquad\qquad \qquad\qquad \alpha_{0,n}^z+\beta_{0,n}^z(A(v)^{-1}-\delta_{0,n}^z)^{-1}\gamma_{0,n}^z\,,
\end{align}
where $v=(v_1,\ldots, v_{r_{n+1}})$, $A(v)= \diag(v_1,\ldots,v_{r_{n+1}})$. 

For fixed $z$ with $\Im z>0$, consider $A(v)$ as a function on\footnote{Here, $\overline{\C}_+$ denotes the closed upper half plane which includes the real line} $\overline{\C}_+^{\, r_{n+1}}$. The matrix
$H_{0,n}-\Phi_n A\Phi_n^*-z$ has an anti-Hermitian part bounded above by
$-(\Im z)\one$, so its inverse exists, is analytic in a neighborhood of $v \in \overline{\C}_+^{\,r_{n+1}}$, and has
norm at most $(\Im z)^{-1}$. Each compressed resolvent entry is therefore
a bounded analytic function of $v \in \C_+^{r_{n+1}}$, continuous on the boundary, and the limit for $|v_j|\to\infty$ exists.
Using the Cauchy-formula, integrating against $[\pi(1+v_j^2)]^{-1}\dd v_j$ evaluates this function at $v_j=i$.
Thus, integrating over all boundary coordinates replaces $A(v)$ by
$i\one_{r_{n+1}}$. Therefore, one obtains
\begin{equation}\label{eq-S_bar_n}
S_{\bar{\m}_n}(z)=\alpha_{0,n}^z+i\beta_{0,n}^z(\one_{r_{n+1}}-i\delta_{0,n}^z)^{-1}\gamma_{0,n}^z\,.
\end{equation}

At each real $\lam\notin\A^1_{0,n}$, the boundary blocks extend
analytically to a neighborhood of $\lam$. Since $\delta^\lam_{0,n}$
is Hermitian, $\one-i\delta^\lam_{0,n}$ is invertible. Thus the displayed
formula for $S_{\bar\m_n}$ extends analytically across that neighborhood and its anti-Hermitian part has a locally uniform boundary limit. 

Applying the Stieltjes inversion formula shows that
$\bar\m_n$ is absolutely continuous there, with density
$\pi^{-1}\Im S_{\bar\m_n}(\lam+i0)$. 
Here, we use the notations
$$
\Im\, M\,=\, \frac{1}{2\imath} \, (M-M^*)\,
$$
for square matrices. 
Consequently, the singular part of $\bar \m_n$
is supported on the finite set $\A^1_{0,n}$.
Since $0,n,\lam$ are fixed parameters for now, we will write
$$
\alpha=\alpha^\lam_{0,n},\quad \beta=\beta^\lam_{0,n},\quad \gamma=\gamma^\lam_{0,n},\quad \delta=\delta^\lam_{0,n},\quad \one=\one_{r_{n+1}}
$$
to simplify notations. By \eqref{eq-S_bar_n} we have for $\lam \not \in \A^1_{0,n}$ that 
\begin{align*}
\Im[S_{\bar{\m}_n}(\lam+i0)]\,&=\,\lim_{\eta \to 0^+} \Im[S_{\bar{\m}_n}(\lam+i\eta)]\,=\,\\
&=\frac{1}{2i}\left[\alpha+i\beta(\one-i\delta)^{-1}\gamma-\alpha^*+i\gamma^*(\one+i\delta^*)^{-1}\beta^*\right]\,.
\end{align*}
Then, since $\alpha$ and $\delta$ are Hermitian matrices, and $\beta=\gamma^*$, the above formula becomes
\begin{align*}
\Im[S_{\bar{\m}_n}(\lam+i0)]\,&=\,\gamma^*\frac{(\one-i\delta)^{-1}+(\one+i\delta)^{-1}}{2}\gamma \\
&=\,\gamma^*(\one+\delta^2)^{-1}\gamma\,=\,\beta(\one+\delta^2)^{-1}\gamma\,.
\end{align*}
To summarize, denoting the explicit dependence on $E$ and $n$,
\begin{equation}\label{eq-bfv}
\Im[S_{\bar{\m}_n}(\lam+i0)]\,=\,|\bfv_n^\lam|^2 \qtx{where} \bfv_n^\lam:=\left(\one_{r_{n+1}}+(\delta^\lam_{0,n})^2\right)^{-1/2}\gamma^\lam_{0,n}\,.
\end{equation}

On the other hand, for $\lam \not \in \A_{0,n}=\A_{0,n}^1\cup \A^2_{0,n}$ observe that for any block matrix 
$$
\bfu_n^\lam:=\pmat{B \\ \delta^\lam_{0,n} B}\in \D^\lam_{n}
$$
and
$$
\bfw_n^\lam:=(\one_{r_{n+1}}+(\delta^\lam_{0,n})^2)^{1/2}B
$$
the following identities hold
\begin{align*}
\big(\bfv_n^\lam\big)^*\bfw_n^\lam&=\beta^\lam_{0,n}\left(\one_{r_{n+1}}+(\delta^\lam_{0,n})^2\right)^{-\sfrac12}\left(\one_{r_{n+1}}+(\delta^\lam_{0,n})^2\right)^{\sfrac12}B=\beta^\lam_{0,n} B=\one_{r_0}\,,
\end{align*}
$$
|\bfu_n^\lam|^2=|B|^2+B^*(\delta^\lam_{0,n})^2B=|\bfw_n^\lam|^2\,.
$$
Therefore, applying the matrix Cauchy-Schwarz inequality (Lemma~\ref{lem:matrix-cs}), we obtain
$$
\big|\bfv_n^\lam\big|^2\geq\big|\bfw_n^\lam\big|^{-2}=\big|\bfu_n^\lam\big|^{-2}\,.
$$
Moreover, equality is achieved by the element which minimizes $|D|^2$ within $\D^\lam_n$.
Indeed, by Proposition~\ref{prop:HSmin}, this is the unique element minimizing the Hilbert Schmidt norm and given by
\begin{align}
D_n^\lam&:=\pmat{\hat{B} \\ \delta^\lam_{0,n} \hat{B}}\,,\quad\text{with}\quad \label{eq-hat-u}\\
 \hat{B}&:=\left(\one_{r_{n+1}}+(\delta^\lam_{0,n})^2\right)^{-1}\gamma^\lam_{0,n}\left(\beta^\lam_{0,n}\left(\one_{r_{n+1}}+(\delta^\lam_{0,n})^2\right)^{-1}\gamma^\lam_{0,n}\right)^{-1}\,.
 \label{eq-hat-B}
\end{align}
Since $\one_{r_{n+1}}+(\delta^\lam_{0,n})^2>\one_{r_{n+1}}$, and $\beta^\lam_{0,n}=(\gamma^\lam_{0,n})^*$ is of full rank, the latter inverse exists. One readily checks that
$$
\left| D^\lam_n \right|^{2}\,=\,\left( \beta^\lam_{0,n}\left(\one_{r_{n+1}}+(\delta^\lam_{0,n})^2 \right)^{-1}\gamma^\lam_{0,n} \right)^{-1}=\big|\bfv_n^\lam\big|^{-2}\,.
$$
Hence,
\begin{equation}\label{eq-bfv-D}
\big|\bfv^\lam_n\big|^2\,=\,
\max\,\left| \D^\lam_{0,n}  \right|^{-2}\,=\, \left| D^\lam_n \right|^{-2}\,.
\end{equation}
In particular, we remark that $\lam \mapsto \max\,\left| \D^\lam_{0,n}  \right|^{-2}$ is measurable.

By the Stieltjes inversion formula, there is a point measure $\boldsymbol{\zeta}_n$ supported on the finite set $\A^1_{0,n}$ such that
\begin{equation}\label{eq-mu_bar}
\bar \m_n(\dd \lam)\,=\, \boldsymbol{\zeta}_n(\dd \lam)\,+\, \frac1\pi |\bfv^\lam_n|^2\,\dd\lam
\end{equation}
where $\dd \lam$ denotes the Lebesgue measure.
It remains to identify the singular part $\boldsymbol{\zeta}_n$. 

For $\lam_0\in\R$, consider
\begin{equation*}
\Hi_{n,\lam_0} :=\left\{\psi\in\C^{\G_{0,n}}: H_{0,n}\psi=\lam_0\psi,\quad \Phi_n^*\psi_n=0
\right\},
\end{equation*}
and let $\Pp_{n,\lam_0}$ denote the orthogonal projection onto this space.

After extension by zero outside $\G_{0,n}$, the vectors in $\Hi_{n,\lam_0}$ are precisely the eigenfunctions of $H$ with eigenvalue $\lam_0$ supported in $\G_{0,n}$. Indeed, the condition $\Phi_n^*\psi_n=0$ makes the coupling to $\Sb_{n+1}$ vanish. Conversely, if an eigenfunction of $H$ is supported in $\G_{0,n}$, its equation on $\Sb_{n+1}$ gives
\begin{equation*}
\Upsilon_{n+1}\Phi_n^*\psi_n=0.
\end{equation*}
Since $\Upsilon_{n+1}$ has full column rank, this implies
$\Phi_n^*\psi_n=0$.

In Lemma~\ref{lem-zeta} below we show that
\begin{equation}\label{eq-zeta-nu}
\boldsymbol{\zeta}_n(\{\lam_0\})=\bar\m_n(\{\lam_0\})=\Upsilon_0^*\Pp_{n,\lam_0}\Upsilon_0=\n_n(\{\lam_0\})\,.
\end{equation}
Note that the first and last equality have already been explained and we need to derive the middle one.
As $E_0$ is arbitrary and both, $\boldsymbol{\zeta}_n$ and $\n_n$ are point measures, this means $\boldsymbol{\zeta}_n=\n_n$
which with \eqref{eq-mu_bar} and the definition \eqref{eq-def-nu_n^perp} gives
\begin{equation}\label{eq-nu^perp-bfv}
\n_n^\perp(\dd \lam)\,=\,\frac1\pi \big| \bfv_n^\lam\big|^2 \,\dd \lam\,.
\end{equation}
Using the identities \eqref{eq-bfv}, \eqref{eq-bfv-D} and Proposition~\ref{prop-1}~(i) the formulas for $\n_n^\perp$ of Theorem~\ref{Theo-av} then follow.
Furthermore, $\n_n=\boldsymbol{\zeta}_n$ is supported on $\A^1_n$. As for $m\geq n$ we have $\n_m \geq \n_n$,  (as $\Hi_n \subset \Hi_m$), the measure $\n_n$ is supported on
$\bigcap_{m\geq n} \A^1_{0,m}$. Therefore, the point measure $\n$ is supported on $\displaystyle\bigcup_{n\geq 0} \bigcap_{m\geq n} \A^1_{0,m}$.
Together with Lemma~\ref{lem-zeta} this finishes the proof of Theorem~\ref{Theo-av}. \hfill $\Box$

\begin{lemma}\label{lem-zeta}
We have that
$$
\bar\m_n(\{\lam_0\})=\Upsilon_0^*\Pp_{n,\lam_0}\Upsilon_0\,.
$$
\end{lemma}
\begin{proof}
Let $A(v)=\diag(v_1,\ldots,v_{r_{n+1}})$ and denote the columns of $\Phi_n$ by $\phi_1,\ldots,\phi_{r_{n+1}}$, viewed as vectors in $\C^{\G_{0,n}}$. Fix all coordinates of $v$ except $v_j$. In this setting, the dependence of $H_{0,n,A(v)}$ on $v_j$ is the rank-one perturbation $-v_j\phi_j\phi_j^*$. Then, scalar spectral averaging at the
singleton $\{\lam_0\}$ gives
\begin{equation*}
\int_{\R} \left\langle \phi_j,\, \mathbf{1}_{\{\lam_0\}}(H_{0,n,A(v)})\phi_j
\right\rangle\,\frac{1}{\pi(1+v_j^2)} \dd v_j=0.
\end{equation*}
The integrand is nonnegative, so it vanishes for almost every $v_j$. Since $\mathbf{1}_{\{\lam_0\}}(H_{0,n,A(v)})$ is an orthogonal
projection, this implies
\begin{equation*}
\mathbf{1}_{\{\lam_0\}}(H_{0,n,A(v)})\phi_j=0\,.
\end{equation*}
Now, by Fubini's theorem, these identities hold simultaneously for all $j$ and for Lebesgue-almost every $v\in\R^{r_{n+1}}$.
Consequently, for almost every $v$, every eigenvector of $H_{0,n,A(v)}$ with eigenvalue $\lam_0$ is orthogonal to the range of $\Phi_n$. For such a vector, the boundary perturbation vanishes
and it is also an eigenvector of $H_{0,n}$, which menas it is in $\Hi_{n,E_0}$, thus, $\ker(H_{0,n,A(v)}-E_0) \subset \Hi_{n,E_0}$ for Lebesgue almost all $v$.
On the other hand, one readily checks $\Hi_{n,E_0} \subset \ker(H_{0,n,A}-E_0)$.
Hence,
\begin{equation*}
\ker(H_{0,n,A(v)}-\lam_0\one)=\Hi_{n,\lam_0}\quad \text{for Lebesgue almost all $v\in\R^{r_{n+1}}$}\,.
\end{equation*}
The product Cauchy distribution is absolutely continuous with respect to the Lebesgue measure and has total mass one. Averaging therefore gives
\begin{align}
\bar\m_n(\{\lam_0\})
&=\int_{\R^{r_{n+1}}}\Upsilon_0^*\mathbf{1}_{\{\lam_0\}}(H_{0,n,A(v)})\Upsilon_0\prod_{j=1}^{r_{n+1}}\frac{\dd v_j}{\pi(1+v_j^2)} \,=\,\Upsilon_0^*\Pp_{n,\lam_0}\Upsilon_0\,.
\end{align}
This finishes the proof.
\end{proof}

\section{Multiplicity criterion for the absolutely continuous spectrum \label{sec-mult}}

In this section we will prove Theorem~\ref{Theo} and Theorem~\ref{Theo-2}.
Let us start with Theorem~\ref{Theo-2}, as the proof is shorter.

\renewcommand{\proofname}{Proof of Theorem~\ref{Theo-2}}

\begin{proof}
For convenience, for each $k$ let
\[
\bff_k(\lam)=\frac{1}{\pi}\max\left|\D^\lam_{n_k}\right|^{-2}\,=\,\frac1\pi \left| D^\lam_{n_k}\right|^{-2}
\]
denote the density of $\n_{n_k}^{\perp}$ from Theorem~\ref{Theo-av}, where
$ D^\lam_{n_k}$ is defined as in \eqref{eq-hat-u}, \eqref{eq-hat-B}.
By Proposition~\ref{prop:HSmin},  $\min\|\D^\lam_n\|$ is also achieved at $D^\lam_n$, meaning
\begin{equation}
\min \| \D^\lam_n\|\,=\,\|D^\lam_n\|\,.
\end{equation}
Now let
$$
C(\lam):=\sup_k\left\|D^\lam_{n_k}\right\|^2<\infty \quad \text{for almost every $\lam$ in $(a,b)$}
$$
and note that $\lam\mapsto C(\lam)$ is measurable.
Then,
\[
\left| D^\lam_{n_k}\right|^2\,\leq\, C(\lam)\one_{r_0},
\qquad
\left|D^\lam_{n_k}\right|^{-2}
\geq C(\lam)^{-1}\one_{r_0},
\]
implying 
\[
\bff_k(\lam)\geq \frac{1}{\pi C(\lam)}\one_{r_0}
\]
for almost every $\lam\in (a,b)$. In particular,
$$
\bar \m_{n_k}(\dd \lam)\,\geq\, \n_{n_k}^\perp (\dd \lam)\,\geq\, \frac{1_{(a,b)}(\lam)}{C(\lam)} \one_{r_0} \,\dd \lam\,.
$$
As finite Borel measures on $\R$ are Radon measures, integrating against positive continuous functions, we see in the limit
$$
\m(\dd \lam)\,\geq\, \n^\perp(\dd \lam)\,\geq\, \frac{1_{(a,b)}(\lam)}{C(\lam)} \one_{r_0}\, \dd \lam\,.
$$
Splitting $\m$ or $\n^\perp$ into its singular and absolutely continuous parts and approximating indicator functions with continuous functions, we see that
$$
\n^\perp_{ac}(\dd \lam)\,\geq\,
\frac{1_{(a,b)}(\lam)}{C(\lam)} \one_{r_0}\,\dd \lam\,.
$$
In particular, if $\n^\perp_{ac}(\dd \lam)=\bff(\lam)\dd \lam$, its density $\bff(\lam)$ is almost surely bounded below by $\frac{1}{C(\lam)} \one_{r_0}$ in $(a,b)$ and hence given by an invertible matrix.
This finishes the proof of Theorem~\ref{Theo-2}.
\end{proof}

Now let us turn to the proof of Theorem~\ref{Theo}.
\renewcommand{\proofname}{Proof of Theorem~\ref{Theo}}
\begin{proof}

By \eqref{eq-str-T} and Proposition~\ref{prop:HSmin} we may select 
$T=T(\lam,n)\in \T^\lam_{n\leftarrow 0}$ 
which minimizes the Hilbert-Schmidt norm, then it will also minimize the usual norm and
$\|T\|=\min \|\T^\lam_{n\leftarrow 0}\|$.
Following the proof of Proposition~\ref{prop:HSmin}, $T=T(E,n)$ is the unique element of $\T^\lam_{n\leftarrow 0}$ such that for any $v\in \ker(\beta^\lam_{0,n})$ we have
\begin{equation} \label{eq-cond-minT}
\pmat{v\\\delta^\lam_{0,n} v}^* T= \nul\,.
\end{equation}
To write down the complex formula without getting lost in the multiple indices, we will consider $\lam$ and $n$ as fixed parameters
and use the short notations like above
$$
\alpha=\alpha^\lam_{0,n},\quad \beta=\beta^\lam_{0,n},\quad \gamma=\gamma^\lam_{0,n},\quad \delta=\delta^\lam_{0,n}\,,\quad \one=\one_{r_{n+1}}\,.
$$
Then, similar to \eqref{eq-hat-u}, \eqref{eq-hat-B}, $T(\lam,n)$ is given by
\begin{align}
&T=T(\lam,n)\,=\,\pmat{D^\lam_n & N^\lam_n} \,=\,\pmat{\hat B & -\hat \bbb \\ \delta \hat B & \gamma-\delta \hat \bbb} 
\qtx{where}\\
&\hat B\,=\,\left(\one+\delta^2\right)^{-1}\gamma\left(\beta\left(\one+\delta^2\right)^{-1}\gamma\right)^{-1} \label{eq-hatB-2}\\
&\hat \bbb\,=\,\left(\one+\delta^2\right)^{-1}\left(\delta \gamma-\gamma\left(\beta\left(\one+\delta^2\right)^{-1}\gamma\right)^{-1} \left(\beta \left(\one+\delta^2\right)^{-1}\delta \gamma-\alpha \right)\,\right).
\end{align}
One readily checks that $\beta \hat B=\one$, $\beta \hat \bbb= \alpha$ and that $\beta v =0$ ($\Leftrightarrow v^*\gamma=0$) implies \eqref{eq-cond-minT}.
Now it is easy to see that
$$
\lam\mapsto \min \|\T^\lam_{n\leftarrow 0} \|=\| T(E,n)\|
$$
is measurable. The hypothesis of the Theorem can be stated as
\begin{equation}\label{eq-hip}
\liminf_{n\to\infty} \int_a^b \|T(E,n)\|^{2p}\, \dd E\,<\, \infty\,.
\end{equation}

Note that equations \eqref{eq-hat-B} and \eqref{eq-hatB-2} are identical, meaning that the matrix $D^\lam_n$ here is indeed the same $D^\lam_n$ as in \eqref{eq-hat-u}.
Thus, by \eqref{eq-hat-u}, \eqref{eq-hat-B}, \eqref{eq-bfv-D}, \eqref{eq-hatB-2} we find
\begin{equation*}
\left|\bfv_n^\lam\right|\,=\,\left|D^\lam_n\right|^{-1}\,=\,\left|T(\lam,n)\pmat{\one_{r_0}\\\nul}\right|^{-1}.
\end{equation*}
The Hermitian symplectic property gives (using $T=T(E,n)$)
\begin{equation*}
\one_{r_0}
=\pmat{\nul &\one_{r_0}}J_{r_0}\pmat{\one_{r_0}\\ \nul}
=\left(T \pmat{\nul \\\one_{r_0}}\right)^*
J_{r_{n+1}}T \pmat{\one_{r_0}\\\nul}=\left(D^\lam_n\right)^* J_{r_{n+1}} N^\lam_n\,.
\end{equation*}
Thus, Lemma~\ref{lem:matrix-cs} implies
\begin{equation*}
|\bfv_n^\lam|^2
=\left| D^\lam_n \right|^{-2}
\leq\left|J_{r_{n+1}} N^\lam_n\right|^2\,\leq\, \norm{N^\lam_n}^2 \one_{r_0}\,\leq\,\norm{T(E,n)}^2 \one_{r_0}\,.
\end{equation*}
For the densities of the a.c. measures $\n_n^\perp$, which we denote by
\begin{equation}\label{eq-fk-T}
\bff_n(\lam):=\frac1\pi\left|\bfv^\lam_n\right|^2\,=\,\frac1\pi\,\left| D^\lam_n \right|^{-2}\,,
\end{equation}
the hypothesis \eqref{eq-hip} of the Theorem yields
\begin{equation*}
\liminf_{n\to\infty}\int_a^b\|\bff_n(\lam)\|^p\,\dd\lam
\leq\frac1{\pi^p}\liminf_{n\to\infty}\int_a^b
\left\|T(E,n)\right\|^{2p}\,\dd\lam<\infty.
\end{equation*}

Using weak compactness in $L^p((a,b),\C^{r_0\times r_0})$ and Banach-Alaoglu, we find a weakly convergent subsequence  $\bff_{n_k}\rightharpoonup \bff\in L^p((a,b),\C^{r_0\times r_0})$ where along that subsequence,
\begin{equation}\label{eq-n_k-T-bd}
\sup_k \int_a^b \|T(\lam,n_k)\|^{2p}\,\dd \lam\,<\,\infty\,.
\end{equation}

The limit is positive semidefinite almost everywhere. Testing against $g\in C_c((a,b))$, extended by zero to $\R$, and using $\n_{n_k}^\perp\rightharpoonup\n^\perp$, we obtain
\begin{equation*}
\n^\perp(g)=\lim_{k\to\infty}\int_a^b g(\lam)\bff_{n_k}(\lam)\,\dd\lam
=\int_a^b g(\lam)\bff(\lam)\,\dd\lam.
\end{equation*}
Thus, $\n^\perp=\bff(\lam)\,\dd\lam$ on $(a,b)$, proving absolute continuity.\\

Now let us see that $\bff(\lam)$ is almost surely invertible in $(a,b)$.
Suppose, for contradiction, that $\rank\bff(\lam)<r_0$ on a measurable set of positive Lebesgue measure. 
By inner regularity, this set contains a compact set $K\subset(a,b)$ of positive Lebesgue measure. 
Now, the sets $K_r=\{\lam\in K\,:\,\rank \bff(\lam)=r\}$ are measurable, and within $K_r$, the function $\lam\mapsto \ker \bff(\lam)$ is measurable, where $\ker \bff(\lam)$ 
is considered as an element of the Grassmann manifold of $(r-r_0)$-dimensional subspaces.
Hence, one may choose a bounded measurable function $w:\R\to\C^{r_0}$, supported in $K$, such that
\begin{equation*}
w(\lam)\in\ker\bff(\lam),\qquad \|w(\lam)\|=1 \qtx{for all} \lam\in K\,.
\end{equation*}
The linear map
\begin{equation*}
T_w:L^p((a,b),\C^{r_0\times r_0})\longrightarrow L^p(a,b),
\qquad T_w\bfh(\lam)=w(\lam)^*\bfh(\lam)w(\lam),
\end{equation*}
is bounded. Consequently, $h_k:=T_w\bff_{n_k}$ converges weakly to $T_w\bff$ in $L^p(a,b)$.
As $w$ is in the kernel of $\bff(\lam)$ for $\lam \in K$, and it is zero outside $K$, we get for every $g\in C_b(\R)$,
\begin{equation*}
(w^*\n^\perp w)(g)\,=\,\int_a^b g(\lam)w(\lam)^*\bff(\lam)w(\lam)\,\dd\lam\,=\,
\lim_{k\to\infty}\int_a^b g(\lam)h_k(\lam)\,\dd\lam
=0.
\end{equation*}

On the other hand, for almost every $\lam\in K$, Lemma~\ref{lem:sigma} gives
\begin{align*}
h_k(\lam)\,=\,w(\lam)^*\bff_{n_k}(\lam)w(\lam)
&\,\geq\,\sigma_{\min}^2\left(\bff_{n_k}(\lam)\right)\,=\,\frac1{\pi\left\| D^\lam_{n_k} \right\|^2}.
\end{align*}
It follows that, almost everywhere on $K$,
\begin{align*}
-\log h_k(\lam)\,
&\leq\,\log\left(\pi\left\|D^\lam_{n_k}\right\|^2\right)
\,\leq\,\log(\pi)+\frac1p \log\left( \left\|D_{n_k}^\lam\right\|^{2p}\right)\\
&\leq\,\log(\pi)+ \left\|D_{n_k}^\lam\right\|^{2p}\,\leq\,\log(\pi)+\left\| T(E,n_k)\right\|^{2p}\,.
\end{align*}
Using \eqref{eq-n_k-T-bd} we find
\begin{equation*}
\liminf_{k\to \infty}\int_K-\log h_k(\lam)\,\dd\lam
\leq \log(\pi) |K|\,+\,\sup_{k}\int_a^b\left\|T(\lam,n_k) \right\|^{2p}\,\dd\lam<+\infty\,.
\end{equation*}
Applying the Deift--Killip criterion (Proposition~\ref{prop-C1}) with scalar weight equal to $1$ gives $(w^*\n^\perp w)(K)>0$. This contradicts $w^*\n^\perp w=0$, so our assumption was false. \\
Hence, $\rank\bff(\lam)=r_0$ for Lebesgue-almost every $\lam\in(a,b)$. Since $\n$ is a point measure, $\bff$ is also the density of $\m_{ac}$ on this interval. Consequently, the density of the absolutely continuous part of $\m$ has Lebesgue-almost-surely full rank $r_0$ on $(a,b)$.
\end{proof}

\renewcommand{\proofname}{Proof}

\section{Infinite multiplicity for random models}\label{ApplicationRM}

In this section, we prove Theorems \ref{application_1} and \ref{application_2}. 
The key estimates required for the existence of the a.c. spectrum of the operators under consideration are those computed in \cite[Section~9]{Sa-Tr} with the sole modification of using a root matrix instead of a root vector.

With this adaptation, we apply our multiplicity criterion to establish these theorems. We begin with our first random model, which describes independent wires coupled by a random potential.

\renewcommand{\proofname}{Proof of Theorem~\ref{application_1}}
\begin{proof}
Let $s_0$ denote the cardinality of the shell $\Sb_0$ and let us choose $\Ups_0=\one_{s_0}$, so that $r_0=s_0$. With this choice, the computations in the proof of \cite[Theorem~5]{Sa-Tr} remain essentially unchanged: the transfer-matrix products have $2s_0$ columns instead of two, which introduces only a fixed finite constant in the matrix-norm estimates. 

Retaining the choices
$$
  \Phi_n=\one_{s_n}\quad\text{and}\quad\Ups_n=\pmat{\one_{s_{n-1}}\\\nul}\quad\text{for}\, n\geq 1\,,
$$ 
the random operator $H^{(1)}_\omega$ is of the form \eqref{eq-H}.
The transfer matrices  given by
$$
  {T}_n^\lam:=
  \pmat{
    (A_n+V_n-\lam\one_{s_n})\Ups_n & -\Ups_n\\
    \Ups_n & \nul},
  \quad\text{with}\quad A_n=\diag(a_1,\ldots,a_{s_n})\,
$$
satisfy ${T}_n^\lam\in \T^\lam_n$ and 
$$
  \liminf_{n\to\infty}\int_a^b\EE\, \norm{{T}_n^\lam\cdots T_0^\lam}^{4}\,\dd\lam<\infty\,.
$$
 By Fubini's theorem and Fatou's lemma this means that for almost all $\omega\in\Omega$
 $$
\liminf_{n\to\infty} \int_a^b \min \norm{\T^\lam_{n\leftarrow 0}}^4\, \dd\lam\,\leq\,  \liminf_{n\to\infty}\int_a^b \norm{{T}_n^\lam\cdots T_0^\lam}^{4}\,\dd\lam<\infty\,.
 $$
Therefore, applying Theorem~\ref{Theo} with $p=2$, we obtain for almost all $\omega$, $\n^\perp$ is almost surely continuous and  that $m_{\ac}(H_\omega^{(1)},\lam)\geq s_0$ for $\lam\in I_1=(-2+\sup_j a_j\,,\, 2+\inf_j a_j)$. 
Moreover, it is easy to check that in this model there can not be compactly supported eigenfunctions, therefore $\n=0$ and the spectral measure at any vector in $\C^{\Sb_0}$ is purely absolutely continuous in $I_1$.

Now, to obtain larger lower bounds, fix $N\geq1$ and consider the tail
\[
  \G_{N,+}:=\bigsqcup_{n\geq N}\Sb_n.
\]
Let $H_{\omega,N}^{(1)}$ be the restriction of $H_\omega^{(1)}$ to
$\ell^2(\G_{N,+})$, with the connection to $\Sb_{N-1}$ removed. After reindexing the shells, this is another stair-like model satisfying the same properties as the original model. Its initial shell is $\Sb_N$, of size $s_N$. Applying the preceding argument with root
matrix $\one_{s_N}$ gives
$$
  m_{\ac}(H_{\omega,N}^{(1)})\geq s_N\,,
$$
in $I_1$,  for almost all $\omega$.  
Moreover, if $H_{\omega,[0,N-1]}^{(1)}$ denotes the restriction of $H^{(1)}_\omega$ to $\ell^2\left(\G_{0,N-1}\right)$ and 
$$
  \wh H_{\omega,N}^{(1)}:=H_{\omega,[0,N-1]}^{(1)}\oplus H_{\omega,N}^{(1)}\,,
$$
then $H^{(1)}_\omega - \wh H^{(1)}_{\omega,N}$ has finite rank. By the Kato--Rosenblum Theorem, the absolutely continuous parts of these two operators are unitarily equivalent.  Since $H_{\omega,[0,N-1]}^{(1)}$ acts on a finite-dimensional space, it has no absolutely continuous part. Hence, for $\lam\in I_1$ and almost all $\omega$,
$$
  m_{\ac}(H_\omega^{(1)},\lam) =m_{\ac}(H_{\omega,N}^{(1)},\lam) \geq s_N
$$
in the interval $I_1$. 
Also, by the same argument as above, the spectral measure of $H_{\omega,N}^{(1)}$ at any vector in $\C^{\Sb_N}$ is purely absolutely continuous in $I_1$, for almost all $\omega$.
These assertions hold for every $N\in\NN$ and all $\omega \in \Omega_N$, a set of measure $1$. 
Consequently, for $\omega \in \tilde \Omega:=\bigcap_{N=1}^\infty \Omega_N$, (a set of full measure $1$) we have
$$
  m_{\ac}(H_\omega^{(1)},\lam)\geq\sup_{N\geq 1}s_N=\infty\,.
$$
Moreover, adopting \cite[Lemma 2.2]{FLSSS}, for $\omega \in \tilde \Omega$,
the spectrum in $I_1$ is almost surely absolutely continuous.
This yields the desired result.
\end{proof}
\renewcommand{\proofname}{Proof}

We now turn to the proof of Theorem~\ref{application_2}, concerning the Bethe lattice. As shown in the proof of \cite[Theorem~6]{Sa-Tr}, a unitary transformation reduces this model to the stair-like model, up to a constant rescaling. The existence of absolutely continuous spectrum and the assertion of infinite multiplicity therefore follow from Theorem~\ref{application_1}.

\renewcommand{\proofname}{Proof of Theorem~\ref{application_2}}
\begin{proof}
As shown in the proof of \cite[Theorem~6]{Sa-Tr}, there exists a unitary operator $U=\bigoplus_{n\geq0}U_n$ on $\ell^2(\G_2)$ satisfying
$$
  U^*\Delta^{(2)}U=\sqrt{2}\,\Delta^{(1)},
$$
where $\Delta^{(1)}$ is the stair-like operator with $s_n=2^n$ and $a_j=0$ for every $j\geq1$. Define
$$
  \widetilde H_\omega:=\frac{1}{\sqrt{2}}U^*H_\omega^{(2)}U=\Delta^{(1)}+\bigoplus_{n\geq0}\widetilde V_n(\omega),
  \quad\text{with}\quad \widetilde V_n(\omega) :=\frac{1}{\sqrt{2}}U_n^*V_n(\omega)U_n.
$$
Since the matrices $U_n$ are deterministic, the Hermitian matrices $\widetilde V_n$ are independent. By unitary invariance of the operator
norm,
\begin{align*}
  \bigl\|\EE(\widetilde V_n)\bigr\|
    &=\frac{1}{\sqrt{2}}\norm{\EE(V_n)},\\
  \EE\left(\bigl\|\widetilde V_n\bigr\|^{2}\right)
    &=\frac12\EE\bigl(\norm{V_n}^{2}\bigr),\\
  \EE\left(\bigl\|\widetilde V_n\bigr\|^{4}\right)
    &=\frac14\EE\left(\bigl\|V_n\bigr\|^{4}\right).
\end{align*}
Thus the transformed potential satisfies \eqref{decaying-1}. Since $s_n=2^n\to\infty$ and the common band is $(-2,2)$, Theorem~\ref{application_1} implies that, almost surely,
$$
  m_{\ac}(\widetilde H_\omega)=\infty
$$
in the interval $(-2,2)$ and the spectrum is purely absolutely continuous. Finally, 
the relation
$$
  H_\omega^{(2)}=U\bigl(\sqrt{2}\,\widetilde H_\omega\bigr)U^*\,,
$$
implies that $m_{\ac}(H_\omega^{(2)})=\infty$, almost surely in $I_2=(-2\sqrt{2}\,,\,2\sqrt{2})$, as claimed.
\end{proof}
\renewcommand{\proofname}{Proof}

\appendix
\section{On matrix-valued measures and the Stieltjes transform}

In this appendix we briefly explain the extension of the Stieltjes approximation to positive matrix-valued measures.
Let $ \Her_+(\C^m)$ denote the set of $m \times m$ positive semidefinite (Hermitian) matrices.

\begin{definition}
Let $X$ be a set and $\Aa$ a $\sigma$-algebra in $X$. A map $\m:\Aa\longrightarrow \Her_+(\C^m)$ is called a $ \Her_+(\C^m)$-valued measure if the following properties hold:
\begin{enumerate}
\item $\m(A)\in \Her_+(\C^m) $, for each $A\in\Aa$;
\item $\m(\emptyset)=0$.
\item If $\lbrace A_j\rbrace_{j\ge0}$ is a countable family of disjoint sets in $\Aa$, then
\begin{equation*}
\m\left(\bigsqcup_{j=0}^\infty A_j\right)=\sum_{j=0}^\infty\m(A_j)\,,
\end{equation*}
where the convergence of the sum is in the usual norm topology of $\C^{m\times m}$.
\end{enumerate}
\end{definition}
Given a $ \Her_+(\C^m)$-valued measure $\m$ and $v\in\C^m$, one can associate a finite scalar, nonnegative real-valued measure $\m_v$ defined by
\begin{equation*}
\m_v(A)=v^*\m(A)v\,,\quad\forall A\in\Aa\,.
\end{equation*}
This fact allows us to define integration with respect to $\m$.

\begin{proposition}
Let $(X,\mathcal{A})$ be a measurable space and $\m$ a $ \Her_+(\C^m)$-valued measure  on $(X,\mathcal{A})$. Let $\mathcal{M}_b(X,\C)$ be the space of bounded, measurable, complex-valued functions on $X$. There exists a unique linear map $\mathcal{M}_b(X,\C)\longrightarrow \C^{m\times m}$, denoted by $f\mapsto\int_Xf\dd\m$, which satisfies the following properties:
\begin{enumerate}[{\rm (i)}]
\item For every $f\in\mathcal{M}_b(X,\C)$ and $v\in\C^m$, we have that
\begin{equation*}
v^*\left(\int_Xf\dd\m\right)v=\int_Xf\dd\m_v\hspace{0.25cm},
\end{equation*}
\item $\int_X\mathbf{1}_A\dd\m=\m(A)$, for every $A\in \mathcal{A}$,
\end{enumerate}
\end{proposition}

\begin{proof}
 Given $f\in\mathcal{M}_b(X,\C)$, let us consider the map $Q_f:\C^m\longrightarrow\mathbb{C}$ defined by
\begin{equation*}
Q_f(v)=\int_Xf\dd\m_v\,.
\end{equation*}
If $f$ is the indicator function of a set $A\in\mathcal{A}$, then  $Q_f$ is a Hermitian quadratic form on $\C^m$ since
\begin{equation*}
Q_f(v)=\m_v(A)=v^*\m(A)v\
\end{equation*}
and $\m(A)$ is a Hermitian matrix. In turn, if $f$ is a simple measurable function, by linearity, $Q_f$ is still a sesquilinear quadratic form.  Consequently, standard approximation arguments show that $Q_f$ is a sesquilinear quadratic form on $\C^m$ for general complex-valued functions $f$. Moreover, by the Riesz representation theorem, for every  $f\in\mathcal{M}_b(X,\C)$, there exists a unique matrix $\mathcal{I}_f\in \C^{m\times m}$ such that
\begin{equation}
\label{i1}
Q_f(v)=v^*\mathcal{I}_fv,\hspace{0.25cm}\forall v\in\C^m\,.
\end{equation}
We then define the integral of a function $f\in\mathcal{M}_b(X,\mathbb{C})$ with respect to the measure $\m$ as follows:
\begin{equation*}
\int_Xf\dd\m:=\mathcal{I}_f\,.
\end{equation*}
The linearity of the map $f\mapsto\mathcal{I}_f$, together with properties (i) and (ii), follow readily from its definition.
\end{proof}

We define the Stieltjes transform of a   $\Her_+(\C^m)$-valued measure  $\m$ on  $(\R,\Bb(\R))$  by
\begin{equation*}
S_{\m}(z)=\int_{\mathbb{R}}\frac{1}{x-z}\dd\m(x)\,.
\end{equation*}
Accordingly, the anti-Hermitian part of $S_{\m}(z)$ is given by
$$
\Im\left[S_{\m}(z)\right]=\int_{\mathbb{R}}\frac{\eta}{(x-\lam)^2+\eta^2}\dd\m(x)\,,
$$
where $z=\lam+i\eta$. In particular, when $\eta>0$, the matrix $\Im\left[S_{\m}(z)\right]$ belongs to $\Her_+(\C^m)$, and therefore
\begin{equation*}
\dd\m^\eta(\lam)=\left[\frac{1}{\pi} \int_{\mathbb{R}}\frac{\eta}{(x-\lam)^2+\eta^2}\dd\m(x)\right]\dd\lam
\end{equation*}
defines a $\Her_+(\C^m)$-valued measure on $(\R,\Bb(\R))$. In this setting, the following property holds.

\begin{proposition}\label{matrix-stieltjes-approximation}
(\textit{The Stieltjes approximation theorem for matrix-valued measures}) Let $\m$ be an $\Her_+(\C^m)$-valued measure on $(\R,\Bb(\R))$. Then
\begin{equation*}
\lim_{\eta \to 0^+}\int_{\R}f(x)\dd\m^\eta(x)=\int_{\R}f(x)\dd\m(x)\,,
\end{equation*}
for every function $f\in C_b(\R)$. Here, the convergence is in the usual norm (or product) topology of $\C^{m\times m}$.
\end{proposition}



\begin{proof}
Given an $\Her_+(\C^m)$-valued measure $\m$ on $(\R,\Bb(\R))$ and $v\in\C^m$, let $\m_v$ be the scalar measure on  $(\R,\Bb(\R))$ defined by
\begin{equation*}
\m_v(A)=v^*\m(A)v,\hspace{0.25cm}\forall A\in\Bb(\R).
\end{equation*} 
Since $\m_v$ is a finite measure, the Stieltjes transform $S_{\m_v}(z)$ is well defined. Furthermore, according to the Stieltjes approximation theorem, the scalar measure
\begin{equation*}
\dd\m^\eta_v(\lam)=\left[\frac{1}{\pi}\int_{\mathbb{R}}\frac{\eta}{(x-\lam)^2+\eta^2}\dd\m_v(x)\right]\dd\lam
\end{equation*} 
converges weakly to  $\m_v$ as $\eta\to 0^+$. On the other hand
\begin{equation*}
\dd\m_v^\eta(\lam)=v^*\left(\left[\frac{1}{\pi}\int_{\mathbb{R}}\frac{\eta}{(x-\lam)^2+\eta^2}\dd\m(x)\right]\dd\lam\right)v=v^*\left(\dd\m^\eta(\lam)\right)v.
\end{equation*} 
Thus, for every $f\in C_b(\R)$ and $v\in \C^m$, we have
\begin{equation}
\lim_{\eta\to 0}v^*\left(\int_{\R}f(x)\dd\m^\eta(x) \right)v=v^*\left(\int_{\R}f(x)\dd\m(x)\right)v\,.
\end{equation}
This equation, together with the polarization identity, yields the desired result. 
\end{proof}

\renewcommand{\proofname}{Proof}

\section{Some Linear Algebra facts}

Let us begin with a small lemma. 

\begin{lemma}\label{lem-A1}
Let $A\in\mathbb{C}^{k\times \ell}$ be a matrix of rank $\ell$ ($k\geq \ell$) and $\chi\in\mathbb{C}^\ell$. Then,
$$
\chi^*\, |A|^{-2}\,\chi\,=\, \frac{1}{\min\limits_{\zeta\in\mathbb{C}^\ell,\,\zeta^*\chi=1} \|A\zeta\|^2}
$$
where, as before, $|A|=\sqrt{A^*A}$.
\end{lemma}
\begin{proof}
To see this, note that $\zeta^*\chi=1$ implies, by the Cauchy--Schwarz inequality, that
$$
\|A\zeta\|^2\, \left(\chi^*\,|A|^{-2}\chi\right)\, =\, \|(|A|\zeta)\|^2\,\|(|A|^{-1}\chi)\|^2\,\geq\, \left|\, (|A| \zeta)^* |A|^{-1} \chi\,\right|^2\,=\,|\zeta^*\chi|^2\,=\,1.
$$
Furthermore, we get equality for $\zeta=|A|^{-2}\chi\,/\, (\chi^*|A|^{-2}\chi) $.
This finishes the proof of the lemma.
\end{proof}

From this we get a matrix-version of a Cauchy-Schwarz inequality we use in the proof of Theorem~\ref{Theo-av}.
Note that for vectors satisfying $v^*w=1$ the Cauchy-Schwarz inequality gives $1\leq \|v\|^2\, \|w\|^2$ which can be rewritten as $(v^*v)^{-1} \leq (w^*w)$ or $|v|^{-2} \leq |w|^2$, where
$|M|=\sqrt{M^*M}$ in general denotes the matrix absolute value of $M$. 

\begin{proposition}\label{lem:matrix-cs}
Let $X,Y\in\C^{k\times\ell}$, $k\geq \ell$, be matrices such that $X^*Y=I$. Then $$|X|^{-2}\leq|Y|^2\;.$$
\end{proposition}
\begin{proof}
Given $u\in\C^{\ell}$, by Lemma~\ref{lem-A1} there exists $\zeta\in\C^\ell$ such that $\zeta^*u=1$ and
\begin{equation*}
u^* |X|^{-2}u=\frac{1}{\lVert X\zeta \rVert^2}\,.
\end{equation*}
Moreover, by hypothesis, we have that $\zeta^* X^*Yu=1$. Thus, applying the Cauchy-Schwarz inequality, we obtain
\begin{equation*}
\lVert X\zeta \rVert\, \lVert Yu \rVert\geq 1\,.
\end{equation*}
Combining this with the previous identity yields
\begin{equation*}
u^* |Y|^2u=u^*Y^*Yu=\lVert Yu\rVert^2\geq\frac{1}{\lVert X\zeta\rVert^2}=u^* |X|^{-2} u\,.
\end{equation*}
Since $u\in\C^\ell$ was arbitrary, the desired inequality follows.
\end{proof}

Next let us give a small estimate which is important in the proof of Theorem~\ref{Theo}.

\begin{lemma}\label{lem:sigma}
Given a complex matrix $A\in\C^{k\times \ell}$, $k\geq \ell$, let $\sigma_{\min}(A)=\sigma_\ell(A)$ denote its smallest singular value. Then
\begin{equation*}
x^*|A|^2x\geq \sigma_{\min}^2(A)\|x\|^2\,.
\end{equation*}
for all $x\in C^\ell$.
\end{lemma}
\begin{proof}  
Let $A=U\Sigma V^*$ be a singular value decomposition of $A$, and let $\sigma_1,\cdots,\sigma_\ell$ denote its singular values. For $x\in\C^\ell$, define $y=V^*x\in \C^\ell$. Since $U\in\C^{k\times k}$ and $V\in\C^{\ell \times \ell}$ are unitary matrices, we have that
\begin{align*}
x^*|A|^2x&=\|Ax\|^2\,=\,\|U\Sigma V^* x\|^2\,=\,\|\Sigma y\|^2\,=\,\sum_{j=1}^{\ell}\sigma_j^2|y_j|^2\\
& \geq \sigma^2_{\min}(A)\sum_{j=1}^{\ell}|y_j|^2 \,=\,\sigma_{\min}^2(A)\|y\|_2^2\,=\,\sigma_{\min}^2(A)\|x\|^2\,,
\end{align*}
as claimed.
\end{proof}

Finally, let us show that for the type of affine spaces we have for transfer matrices and the Dirichlet data, the element which minimizes the Hilbert-Schmidt norm also minimizes the squared matrix absolute value and the norm.

\begin{proposition}\label{prop:HSmin}
Let $\T\subset \C^{k\times \ell}$ be an affine space of the form $\T=T+\VV^{\oplus \ell}$, where $\VV\subset \C^k$ is a vector subspace and
$$
\VV^{\oplus \ell}=\{[v_1,\ldots,v_\ell]\,:\, v_j \in \VV\}\,\subset\, \C^{k\times \ell}
$$
is the set of $k\times \ell$ matrices whose column vectors belong to $\VV$. \\
Let $A\in \T$ be the matrix which minimizes the Hilbert-Schmidt norm $\|M\|_2=\sqrt{\Tr(M^*M)}$. Then, $A$ also minimizes the squared matrix absolute value $|M|^2=M^*M$ and the usual norm, which means
$$
\forall T\in\T\,:\, |T|^2 \geq |A|^2 \qtx{and} \|T\| \geq \|A\|\,.
$$
\end{proposition}

\vspace{.1cm}

\begin{rem*}
By the Löwner-Heinz Theorem, $|T|^2\geq |A|^2$ also implies $|T|\geq |A|$. Thus, $A$ also minimizes the matrix absolute value within $\T$.
However, we keep the square in the statement as this is what we need in the proofs of the paper. $|T|\geq |A|$ does in general not imply  $|T|^2 \geq |A|^2$,  as the square in general is not an operator-monotone function.
Hence for an arbitrary matrix set $\M$, a minimizer of $|M|$ does not automatically minimize $|M|^2$. It is possible that in terms of the order of positive semidefinite matrices, 
a minimum of the matrices $\{|M|:M\in\M\}$ exists but a minimum of $\{ |M|^2\,:\,M\in\M\}$ does not, e.g.
$\displaystyle \M=\left\{ \pmat{1&0\\0&0}, \pmat{2&1\\1&1} \right\}$.

\end{rem*}
\begin{proof}
Let $V\in \C^{k\times s}$ be a matrix whose column vectors form a basis of $\VV$. Then,
$$
\T\,=\,\{ T+VM\,:\, M \in \C^{s\times \ell} \}
$$
As the Hilbert-Schmidt norm stems from an inner product, the unique element which minimizes the Hilbert-Schmidt norm is given by the orthogonal projection of the zero matrix onto this affine space.
This means it is given by $A\in \T$ which is orthogonal to $\VV^{\oplus \ell}$, this means
$$
0\,=\,\Tr((VM)^*A)\,=\,\Tr(M^* V^*A)\quad\text{ for all $M\in \C^{s\times \ell}$}\,.
$$
But this is equivalent to 
$$
V^*A=\nul \,,\qtx{or,} A^*V=\nul\,.
$$
As $\T$ is an affine space, we have $\T=\{A+VM\,:\,M\in \C^{s\times \ell} \}$. Now, for $T=A+VM \in \T$ we find
$$
|T|^2=T^*T = (A+VM)^*(A+VM)=A^*A+M^*V^*VM\,\geq\, A^*A=|A|^2\,,
$$
which also implies
$$
\|T\|^2=\|\, |T|^2\, \|\,\geq\, \|\, |A|^2\,\|\,=\,\|A\|^2\,.
$$
\end{proof}

\section{Existence of a.c. component in weak limits of measures}

For the reader's convenience, in this section we state the Deift-Killip criterion for existence of absolutely continuous spectrum from \cite{DK}.
We are using Proposition~\ref{prop-C1} in a slightly different manner to get the multiplicity estimate for the a.c. spectrum.

\begin{proposition}\label{prop-C1}
Let $\Omega$ be an open subset of $\R$ and $w(\lam)\in L^1_{loc}(\Omega)$ which is Lebesgue almost everywhere positive. Let $\mu_n$ be a sequence of positive measures which converge weakly to $\mu$ and let $\frac{\dd\mu_n}{\dd\lam}$ denote the Radon-Nikodym derivative with respect to the Lebesgue measure.
Furthermore let $K\subset \Omega$ be a compact set of positive Lebesgue measure and denote
$w(K):=\int_K w(\lam) \dd\lam$. If we have
$$
\liminf_{n\to \infty} \frac{1}{w(K)} \int_K - \log\left( \frac{\dd\mu_n}{\dd\lam} \,\frac{1}{w(\lam)}\right)\,w(\lam)\,\dd\lam\,<\, \infty,
$$
then $\mu(K)>0$. 
\end{proposition}

\begin{rem*}
In particular, if one has this property for any compact set $K\subset \Omega$ of positive Lebesgue measure, then the measure $\mu$ contains an absolutely continuous part in all of $\Omega$, i.e.
$\Omega \subset \supp (\mu_{ac})$, where $\mu_{ac}$ is the absolutely continuous part of $\mu$.
\end{rem*}

This follows directly from the following lemma.
\begin{lemma}[Lemma~2 of \cite{DK}]
Under the assumptions as in Proposition~\ref{prop-C1} we have
$$
\liminf_{n\to \infty} \frac{1}{w(K)} \int_K - \log\left( \frac{\dd\mu_n}{\dd\lam} \,\frac{1}{w(\lam)}\right)\,w(\lam)\,\dd\lam\,\geq \log\left( \frac{w(K)}{\mu(K)}\right)
$$
where $w(K):=\int_K w(\lam) \dd\lam > 0$.
\end{lemma}
\begin{proof}
Let $\Phi$ denote the set of positive continuous functions of compact support which take value 1 on $K$, then
\begin{align*}
\mu(K) = \inf_{\phi\in\Phi} \int \phi \dd\mu &= \inf_{\phi\in\Phi} \lim_{n\to\infty} \int \phi \dd\mu_n \geq
\inf_{\phi\in\Phi} \limsup_{n\to\infty} \int \phi \frac{\dd\mu_n(\lam)}{\dd\lam}\,\dd\lam\\
 &\geq \limsup_{n\to\infty} \int_K \frac{\dd\mu_n(\lam)}{\dd\lam}\,\frac{1}{w(\lam)} \;w(\lam)\;\dd\lam\;.
\end{align*}
Now divide by $w(K)>0$, take   $-\log(\cdot)$ on both sides and use Jensen's  inequality to get the result. Note that taking $-\log(\cdot)$ changes the inequality sign and $\limsup$ to $\liminf$.
\end{proof}

\section*{Acknowledgments}

The authors have received funding from the Chilean Agencia Nacional de Investigación y Desarrollo
with the grants FONDECYT 1230949 , FONDECYT 1252105 and the Scholarship Doctorado Nacional 2022/ folio 21220518.\\
The author C. Sadel thanks the Friedrich-Alexander-Universität Erlangen-Nürnberg for the opportunity to serve as a replacement professor during the winter semester 2025/26.
\setcounter{tocdepth}{1}

\vspace{.3cm}

\subsection*{Data availability}
No data sets were generated or analyzed during the current study.


\vspace{.3cm}

\subsection*{Use of generative AI}
ChatGPT version Astra 6 was used in the production of images and to correct typos. No AI tools were used to produce mathematical results, or to write the manuscript.


\end{document}